\documentclass[journal,11pt,draftcls,onecolumn]{IEEEtran}

\usepackage{cite}

\usepackage{amssymb}
\usepackage[dvips]{graphicx}
\usepackage{psfrag}
\usepackage{subfigure}

\ifCLASSINFOpdf
\else
\fi
\usepackage[cmex10]{amsmath}
\makeatletter
\newlength \figwidth
\if@twocolumn
\else
\fi
\makeatother

\begin{document}

\title{Non-Linear Transformations of Gaussians and Gaussian-Mixtures with implications on Estimation and Information Theory}

\author{Paolo~Banelli,~\IEEEmembership{Member,~IEEE}
\thanks{The author is with the Department of Electronic and Information
Engineering, University of Perugia, 06125 Perugia, Italy (e-mail:
paolo.banelli@diei.unipg.it).}}


%
%

\date{\today}
\maketitle


%


\newtheorem{theorem}{Theorem}
\newtheorem{lemma}{Lemma}
\newtheorem{conjecture}{Conjecture}
\newtheorem{example}{Example}

\begin{abstract}
 This paper investigates the statistical properties of non-linear trasformations (NLT) of random variables, in order to establish useful tools for estimation and information theory. Specifically, the paper focuses on linear regression analysis of the NLT output and derives sufficient general conditions to establish when the input-output regression coefficient is equal to the \emph{partial} regression coefficient of the output with respect to a (additive) part of the input.
A special case is represented by zero-mean Gaussian inputs, obtained as the sum of other zero-mean Gaussian random variables. The paper shows how this property can be generalized to the regression coefficient of non-linear transformations of Gaussian-mixtures. Due to its generality, and the wide use of Gaussians and Gaussian-mixtures to statistically model several phenomena, this theoretical framework can find applications in multiple disciplines, such as communication, estimation, and information theory, when part of the nonlinear transformation input is the quantity of interest and the other part is the noise.
 In particular, the paper shows how the said properties can be exploited to simplify closed-form computation of the signal-to-noise ratio (SNR), the estimation mean-squared error (MSE), and bounds on the mutual information in additive non-Gaussian (possibly non-linear) channels, also establishing relationships among them.
\end{abstract}


%
\IEEEpeerreviewmaketitle

\section{INTRODUCTION}
\label{sec:introduction}
\IEEEPARstart{N}{on-linear} transformations (NLT) of Gaussian random variables, and
processes, is a classical subject of probability theory, with particular
emphasis in communication systems. Several results are available in the
literature to statistically characterize the non-linear transformation
output, for both real \cite{Bussgang:1952, Baum:1957, Price:1958, Davenport:1958,
Blachman:1968, Van:1966, Baum:1969, Levine:1973} and complex \cite{Minkoff:1985,
Banelli:2000, Dardari:2000} Gaussian-distributed input processes.

If the input to the non-linear transformation is the sum of two, or more, Gaussian
random variables, then the overall input is still Gaussian and,
consequently, the statistical characterization can still exploit the wide
classical literature on the subject. For instance, a key point is to
establish the equivalent input-output linear-gain [or linear regression
coefficient (LRC)] of the non linearity. Anyway, if the interest is to infer
only a part of the input by the overall output, and to establish a \emph{partial} LRC (or linear-gain) with respect to this part of the input, it is necessary to compute multiple-folded integrals involving the
non-linear transformation. This task is in general tedious and, sometimes,
also prohibitive.

This paper observes that, if the NLT input is the sum of zero-mean, independent, Gaussian random variables, all the \emph{partial} LRCs are identical, and equal to the \emph{overall} input-output LRC.
This observation, which can also be derived as a special case of the Bussgang Theorem \cite{Bussgang:1952}, highly simplifies the computation of the partial linear-gain, which can be performed by a single-folded integral over the Gaussian probability density function $(pdf)$ of the overall input.
Furthermore, this  property, which holds true also in other cases not covered by the Bussgang Theorem, lets to simplify the computation of the \textit{partial} linear-gain, also when the non-linearity input is the sum of Gaussian-mixtures \cite{Vaseghi:1}.
Gaussian-mixtures are widely used in multiple disciplines, such as to model electromagnetic interference \cite{Middleton:1972}, images background noise \cite{Lee:2005}, financial assets returns \cite{Buckley:2008}, and, more generally, to statistically model clustered data sets.
Actually, it is the similarity of the theoretical results for suboptimal estimators of Gaussian sources impaired by a Gaussian-mixture (impulsive) noise in \cite{Zhidkov:2006}, with those of non-linear transformations of Gaussian random variables in \cite{Rowe:1982}, \cite{Dardari:2000}, \cite{Banelli:2000}, that led to conjecture the existence of the theorems and lemmas analyzed in this paper.

Inspired by those similarities, this papers establishes theoretical links among NLT statistical analysis and estimation theory, in a general framework where the NLT may either represent non-ideal hardware in a communication system (such as amplifiers, A/D converters, etc.) or the non-linear estimator of the information.
In particular, closed-form computation of classical performance metrics such as the signal-to-noise ratio (SNR), the mean-squared error (MSE) of a non-linear estimator, and bounds on the mutual information in additive non-Gaussian (possibly non-linear) channels can be easily derived when a part of the NLT input is the information of interest, and the other part is the noise (or the interference).

The paper is organized as follows. Section \ref{sec:linear_regression} shortly summarizes LRA for NLT and establishes a condition for the equality of the input-output LRC and the LRC of the output $Z=g(Y)$ with respect to another random variable $X$. Section \ref{sec:NL&Sum} establishes \emph{equal-gain} (i.e., equal-LRC) theorems when $Y=X+N$. Section \ref{sec:MixtureExtension} extends the LRC analysis to Gaussian-mixtures. Section \ref{sec:Inf&Est-Theory} is the main contribution of the paper where implication to SNR, MSE and mutual information analysis is highlighted, while conclusions are drawn in the last Section. Appendices are dedicated to proof theorem and lemmas, and also to highlight other examples where the \emph{equal-gain} theorems hold true.
Throughout the paper $G(\cdot;\sigma^2)$ is used to indicate a zero-mean Gaussian \emph{pdf}, $E\{\cdot\}$ is used for statistical expectation, interchangeably with $E_{X_1 \ldots X_N}\{\cdot\}$, which is used, when necessary, to highlight the (joint) \emph{pdf} $f_{X_1,\ldots,X_N}(\cdot)$ involved in the expectation integral.

\section{LINEAR REGRESSION FOR NON LINEAR TRANSFORMATIONS}
\label{sec:linear_regression}
%
%

Lets indicate with $Z=g(Y)$ the NLT of a random variable $Y$.
For any $Y$ and any $g(\cdot)$, the output random variable $Z$ can be
decomposed as a scaled version of the input $Y$ plus an
uncorrelated distortion term $W_y$, as expressed by
\begin{equation}
\label{eq:Z=KyY}
Z=g(Y)=k_y Y+W_y,
\end{equation}
where
\begin{equation}
\label{eq:k_y defin.}
k_y =\frac{E\{ZY\}}{E\{Y^2\}}=\frac{E_Y\{g(Y)Y\}}{E\{Y^2\}}
\end{equation}
is the input-output linear gain (or LRC) that grants the
orthogonality between $Y$ and $W_y$, i.e., $E\{YW_y \}=0$.
By defining the LRC with respect to another random variable $X$, as expressed by
\begin{equation}
\label{Def:k_x}
k_x =\frac{E\{ZX\}}{E\{X^2\}},
\end{equation}
the linear regression model of $Z$ with respect to $X$ would be expressed by
\begin{equation}
\label{Z:kxX}
Z=k_x X+W_x ,
\end{equation}
where $E\{XW_x \}=0$.
For reasons that will be clarified in the next sections, it may be interesting to establish
 when the two LRCs are the same,
as expressed by $k_y=k_x$. To this end, the following Theorem holds true
\begin{theorem}
\label{Th:LinearExpectedValue}
Th:LinearExpectedValue\footnote{The author is in debt with Prof. G. Moustakides for suggesting the existence of this Theorem, and its use to easily prove Theorem \ref{Th:EqGain-Gauss}.}
\textit{If $X$ and $Y$ are two random variables, $g(\cdot )$ is any non-linear single-valued function, and
\begin{equation}
\label{E_X|Y}
E_{X|Y}\{X\}=\alpha y, \qquad  \textrm{with} \qquad  \alpha = \frac{E\{X^2\}}{E\{Y^2\}}
\end{equation}
then}
\begin{equation}
\label{k_y=k_x}
   k_y=\frac{E\{g(Y)Y\}}{E\{Y^2\}}=\frac{E\{g(Y)X\}}{E\{X^2\}}= k_x.
\end{equation}
\end{theorem}
\begin{IEEEproof}
Observing that
\begin{equation}
\label{E_XY}
 E_{XY}\{g(Y)X\} = E_Y\{g(Y)E_{X|Y}\{X\}\},
\end{equation}
equation \eqref{k_y=k_x} immediately follows by direct substitution of \eqref{E_X|Y} in \eqref{E_XY}.
\end{IEEEproof}

Note that the sufficient condition in \eqref{E_X|Y} corresponds to identify when the Bayesian MMSE estimator \cite{Kay:1993} of $X$ is linear (with a proper $\alpha$) in the (conditional) observation $Y=y$ \footnote{Statistical conditions that grants linearity of the MMSE estimator for a generic $\alpha$ are explored in Appendix \ref{App:EquivalenceTheorem}.}.

Another remark is about the computation of $k_x$, which involves a double-folded integral over the \emph{pdf} of $X$ and $Y$. When Theorem \ref{Th:LinearExpectedValue} holds true, this complexity can be significantly reduced by computing $k_y$, which only requests a single-folded integral over the marginal $pdf$ of $Y$.

\section{NLT OF THE SUM OF RANDOM VARIABLES}
\label{sec:NL&Sum}
The general result in Theorem \ref{Th:LinearExpectedValue}, can be specialized to the case of interest in this paper, which focuses on a NLT $g(\cdot)$ that operates on the sum of two independent random variables, i.e., when the two random variables $X$ and $Y$ are linked by a linear model, as expressed by $Y=X+N$.
By means of \eqref{Def:k_x}, in this case it is possible to represent the NLT output as a linear regression with respect to either the \emph{partial} input $X$, or $N$, as expressed by
\begin{equation}
\label{eq:Z=G(X+N)=KxX+Wx=KnN+Wn}
Z=g(X+N)=k_x X+W_x=k_n N+W_n ,
\end{equation}
where
\begin{equation}
\label{eq:k_x and k_n}
k_x =\frac{E_{XN} \{g(X+N)X\}}{P_X }, \qquad \qquad k_n =\frac{E_{XN} \{g(X+N)N\}}{P_N }
\end{equation}
and $P_X=E\{X^2\}$, $E\{XW_x \}=E\{NW_n \}=0$.
In the most general case, the relationship between the three regression coefficients $k_y $, $k_x $, and $k_n $, is summarized by
\begin{equation}
\label{eq:ky-kx-kn general}
\begin{array}{ll}
 P_Y k_y &=E_{XN} \{g(X+N)(X+N)\} \\
 &=E_{XN} \{g(X+N)X\}+E_{XN} \{g(X+N)N\} \\
 &=P_X k_x +P_N k_n , \\
 \end{array}
\end{equation}
which highlights that the linear gain of the overall input is a weighted
sum of the linear gains of each input component, as expressed by
\begin{equation}
\label{eq:ky-kx-kn general2}
k_y =\frac{P_X }{P_X +P_N +2E\{XN\}}k_x +\frac{P_N }{P_X +P_N +2E\{XN\}}k_n.
\end{equation}
Note that, for special cases when $k_x=k_n$, and $X$, $N$ are orthogonal, i.e., $E\{XN\}=0$, then \eqref{eq:ky-kx-kn general2} induces also $k_y =k_x =k_n $.


\subsection{Equal-Gain Theorems}
\label{subsec:EG-Theorems}
This subsection is dedicated to investigate when the LRCs in \eqref{eq:k_y defin.} and \eqref{eq:k_x and k_n} are identical, for random variables $Y=X+N$.
If $\mathcal{F}\{\cdot\}$ is the Fourier transform operator, and  $C_X(u)=E\{e^{j2\pi Xu}\}=\mathcal{F}^{-1}\{f_X(x)\}$ is the characteristic function of $X$, for $Y=X+N$ Appendix \ref{App:EquivalenceTheorem} proves that Theorem \ref{Th:LinearExpectedValue} is equivalent to the following theorem
\vspace{5mm}
\begin{theorem}
\label{Th:ChFunctions}
\textit{If $Y=X+N$, $X$ and $N$ are two independent random variables, and
\begin{equation}
\label{eq:CharFunc}
C_X^{1-\alpha}(u) = C_N^{\alpha}(u), \qquad  \textrm{with} \qquad  \alpha = \frac{E\{X^2\}}{ E\{Y^2\}}
\end{equation}
then, for any non-linear function $g(\cdot )$ in \eqref{eq:k_y defin.}, \eqref{eq:k_x and k_n}}
\begin{equation}
\label{eq:k_y=k_x=k_n}
   k_y=k_x=k_n.
\end{equation}
\end{theorem}
\begin{IEEEproof}
Theorem \ref{Th:CharacteristicFunctions} in Appendix \ref{App:EquivalenceTheorem} establishes that left-hand-side of \eqref{eq:CharFunc} is equivalent to $E_{X|Y}\{\alpha y\}$, which by Theorem \ref{Th:LinearExpectedValue} concludes the proof.
\end{IEEEproof}
\vspace{5mm}

As detailed in Appendix \ref{App:EquivalenceTheorem}, it is not straightforward to verify all the situations when \eqref{eq:CharFunc} holds true. An important scenario where $k_y =k_x =k_n$ is summarized by the following Theorem \ref{Th:EqGain-Gauss}

\vspace{5 mm}
\begin{theorem}
\label{Th:EqGain-Gauss}
\textit{If $X$ and $N$ are zero-mean Gaussian and independent, $Y=X+N$, $g(\cdot )$ any non-linear single-valued function, then property \eqref{eq:k_y=k_x=k_n} holds true.}
\end{theorem}
\vspace{5mm}
\begin{IEEEproof}
By well known properties of Gaussian random variables \cite{Papoulis:1991}, $Y=X+N$ and $X$ are jointly (zero-mean) Gaussian random variables, and consequently the MMSE estimator of $X$ is linear \cite{Kay:1993}, as expressed by
\begin{equation}
\label{eq:E[X|Y]}
   E\{X|Y\}=\frac{E\{XY\}}{E\{Y^2\}}y.
\end{equation}
Furthermore, $E\{XY\}=E\{X(X+N)\}=E\{X^2\}$, which plugged in \eqref{eq:E[X|Y]} concludes the proof by Theorem \ref{Th:LinearExpectedValue}. Alternative proofs can be found in Appendix \ref{App:Bussgang} by exploiting the Bussgang theorem \cite{Bussgang:1952}, and in Appendix \ref{App:EquivalenceTheorem} by exploiting \eqref{eq:CharFunc}.
\end{IEEEproof}
\vspace{5 mm}
%

In general, by equations \eqref{eq:Z=KyY} and \eqref{eq:Z=G(X+N)=KxX+Wx=KnN+Wn},
it is possible to observe that,
\begin{equation}
\label{eq16}
\begin{array}{ll}
 E\{W_y X\}&=E\{(Z-k_y (X+N))X\} \\
 &=k_x E\{X^2\}+E\{W_x X\}-k_y E\{X^2\}-k_y E\{NX\} \\
 &=(k_x -k_y )P_X , \\
 \end{array}
\end{equation}
and analogously $E\{W_y N\}=(k_n -k_y )P_N$.
Due to the fact that in the derivations of \eqref{eq16} it is only
necessary to assume $X$, $N$ to be orthogonal (i.e., $E\{NX\}=0$), and not
necessarily Gaussian, it is demonstrated the following more general theorem

\vspace{5 mm}
\begin{theorem}
\label{Th:Orthogonality}
\textit{If $X$ and $N$ are two orthogonal random variables, $Y=X+N, \; g(\cdot )$ is any single-valued regular function, by the definitions (\ref{eq:Z=KyY}), (\ref{eq:Z=G(X+N)=KxX+Wx=KnN+Wn})
}
\begin{equation}
\label{eq:E{W_yX}=E{W_yN}=0}
E\{W_y X\}=E\{W_y N\}=0  \qquad \textit{iff} \qquad k_y =k_x =k_n .
\end{equation}
\end{theorem}

\vspace{5 mm}
The property $E\{W_y X\}=E\{W_y N\}=0$ in Theorem \ref{Th:Orthogonality}, highlights the key
element that distinguishes independent zero-mean Gaussian random inputs,
with respect to the general situation, when $X$ and $N$ are characterized by
arbitrary $pdf\mbox{s}$. Indeed, for zero-mean Gaussian inputs, by means of
Theorem \ref{Th:EqGain-Gauss} and the sufficient condition in Theorem \ref{Th:Orthogonality}, the distortion term $W_y$ is orthogonal to both the input components $X$ and $N$, while in
general it is orthogonal only to their sum $Y=X+N$. This means that, in the
general case, it is only possible to state that
\begin{equation}
\label{eq:E{W_yX}=-E{W_yN}}
E\{W_y X\}=-E\{W_y N\}\ne 0,
\end{equation}
which is equivalent to link the tree linear gains by (\ref{eq:ky-kx-kn general2}), rather than by the special case in \eqref{eq:k_y=k_x=k_n}.

Another special case is summarized in the following
\vspace{5 mm}
\begin{theorem}
\label{Th:EqGain-SamePDF}
If $X$ and $N$ are two independent zero-mean random variables with identical probability density functions $f_X (\cdot )=f_N (\cdot )$, $Y=X+N$, $g(\cdot )$ is any single-valued regular function, then \eqref{eq:k_y=k_x=k_n} holds true.

\end{theorem}
\begin{IEEEproof}
By observing the definitions of $k_x $ and $k_n $ in \eqref{eq:k_x and k_n}
, it is straightforward to conclude that $k_x =k_n $, when $f_X (\cdot )$ is
\textit{identical} to $f_N (\cdot )$ (note that also $\sigma _X^2 =\sigma _N^2 )$ and,
consequently, due to $E\{XN\}=E\{X\}E\{N\}=0$, \eqref{eq:k_y=k_x=k_n} follows from \eqref{eq:ky-kx-kn general2}. An alternative proof that exploits \eqref{eq:CharFunc}, can be found in  Appendix \ref{App:EquivalenceTheorem}, together with the extension to the sum of $Q$ i.i.d. random variables.
\end{IEEEproof}

\vspace{5 mm}

\subsection{A Simple Interpretation}
\label{subsec:SimpleInterpration}
An intuitive interpretation of the cases summarized by Theorems \ref{Th:ChFunctions}-\ref{Th:EqGain-SamePDF} is that the non-linear function $g(\cdot )$ statistically handles each input component in the same way, in the sense that it does not privilege or penalize any of the two, with respect to the uncorrelated
distortion. In order to clarify this intuitive statement, lets assume that
$X$ and $N$ are zero-mean and uncorrelated, i.e., $E\{XN\}=0$, $g(\cdot )$ is
an odd function, i.e., $g(y)=g(-y)$, and that the goal is to linearly infer
either $X$, or $N$, or their sum $Y=X+N$, from the observation $Z$. Obviously,
in this simplified set-up, also $Z$ is zero-mean, and consequently the
best (in the MMSE sense) linear estimators of, $X$, $N$, and $Y$ are
expressed by \cite{Kay:1993}
\begin{equation}
\label{eq21}
\hat {X}_{\textrm{mmse}}(Z)=\frac{\sigma _X }{\sigma _Z }\rho _{XZ} Z=k_x \frac{\sigma _X^2
}{\sigma _Z^2 }Z,
\end{equation}
\begin{equation}
\label{eq22}
\hat {N}_{\textrm{mmse}}(Z)=\frac{\sigma _N }{\sigma _Z }\rho _{NZ} Z=k_n \frac{\sigma _N^2
}{\sigma _Z^2 }Z,
\end{equation}
\begin{equation}
\label{eq23}
\hat {Y}_{\textrm{mmse}}(Z)=\frac{\sigma _Y }{\sigma _Z }\rho _{YZ} Z=k_y \frac{\sigma _X^2
+\sigma _N^2 }{\sigma _Z^2 }Z=\hat {X}_{\textrm{mmse}}(Z)+\hat {N}_{\textrm{mmse}}(Z),
\end{equation}
where $\rho _{XZ} =E\{XZ\}/\sigma _Y \sigma _Z $, $\rho _{NZ} $, and $\rho
_{YZ} $ are the cross-correlation coefficients for zero-mean random
variables. Note that, as well known \cite{Kay:1993}, the equality $\hat {Y}(Z)=\hat {X}(Z)+\hat {N}(Z)$ in (\ref{eq23}) holds true also
when $k_y \ne k_x \ne k_n $.
Equations (\ref{eq21})-(\ref{eq23}) highlight that, if the two zero-mean inputs $X$ and $N$
equally contribute to the input in the average power sense, i.e., when
$\sigma _X^2 =\sigma _N^2 $, and their non-Gaussian, and non-identical
$pdf\mbox{s}$ $f_X(x)$, and $f_N(n)$, induce $k_x >k_n$ (or $k_x <k_n$),
then $X$ (or $N$) appears less undistorted in the output $Z$ and,
consequently, it gives an higher contribution to the estimation of the sum,
by $\hat{X}$ (or $\hat{N}$).

\section{GENERALIZATION TO GAUSSIAN-MIXTURES}
\label{sec:MixtureExtension}
Due to the fact that the theorems derived so far mostly established sufficient, but not necessary,  conditions for equal-gain, this section first describes a possible way to test if the property in \eqref{eq:k_y=k_x=k_n} may hold true, or not, with respect to a wider class of $pdf\textrm{s}$. Furthermore, the results that are obtained are instrumental to establish inference and information theoretic insights, when random variables are distributed according to Gaussian-mixtures, as detailed in the next section. To this end, lets start from  a situation we are particularly interested to, when $X$ is Gaussian distributed and $N$ is a zero-mean Gaussian-mixture, as expressed by
\begin{equation}
\label{eq:f_N(n) Mixtures}
f_N (n)=\sum\limits_{l=0}^L {\beta _l G(n;\sigma _{N,l}^2)} =\sum\limits_{l=0}^L
{\frac{\beta_l}{\sqrt {2\pi \sigma _{N,l}^2}}
e^{-\frac{n^2}{2\sigma _{N,l}^2 }}} ,
\end{equation}
where $\sigma _N^2 =\sum\limits_{l=0}^L {\beta _l \sigma _{N,l}^2 } $ is the
variance, and $\sum\limits_{l=0}^L {\beta _l } =1$, i.e., $\beta _l \ge 0$
are the probability-masses associated to a discrete random variable, in order
to grant that $f_N (n)$ is a proper $pdf$ with unitary area.
A Gaussian-mixture, by a proper choice of $L$ and $\beta _l $, can accurately fit a wide class of symmetric, zero-mean $pdf$s, and represents a flexible way to test what happens when $N$ departs from a Gaussian distribution. For instance, this quite general framework includes an impulsive
noise $N$ characterized by the Middleton's Class-A canonical model
\cite{Middleton:1972}, where $L=\infty $, $\beta_l=e^{-A}\frac{A^l}{l!}$ are Poisson-distributed weights, $\sigma _{N,l}^2=\frac{l/A+\Gamma }{1+\Gamma }\sigma _N^2 $, and $A$ and $\Gamma$ are the
canonical parameters that control the impulsiveness of the noise \cite{Berry:1981}. Conversely, observe that when $L=0$, and $\beta _0 =1$, the hypotheses of Theorem \ref{Th:EqGain-Gauss} hold true, and consequently \eqref{eq:k_y=k_x=k_n} is verified.

If $X$ and $N$ are independent, $Y=X+N$ is also distributed as a Gaussian-mixture, as expressed by
\begin{equation}
\label{eq:f_Y(y) Mixtures}
\begin{array}{ll}
 f_Y (y)&=f_N (y)\ast f_X (y)\\
  &=\sum\limits_{l=0}^L {\beta _l G(y;\sigma _{N,l}^2)} \ast  G(y;\sigma _{X}^2) = \sum\limits_{l=0}^L {\beta _l G(y;\sigma _{Y,l}^2 )},\\
 \end{array}
\end{equation}
due to the fact that the convolution of two zero-mean Gaussian functions,
still produces a zero-mean Gaussian function, with variance equal to $\sigma _{Y,l}^2 =\sigma _X^2 +\sigma
_{N,l}^2 $. Thus, the LRC $k_y$ can be expressed
by
\begin{equation}
\label{eq:k_y Mixt.Def.}
k_y =\frac{E_Y \{g(Y)Y\}}{\sigma _Y^2 }=\frac{1}{\sigma _Y^2
}\sum\limits_{l=0}^L {\beta _l E_{Y_l } \{g(Y)Y\}} ,
\end{equation}
where $Y_l =X+N_l $ stands for the $l\mbox{-th}$ ``\textit{virtual''} Gaussian random variable
that is possible to associate to the $l\mbox{-th}$ Gaussian $pdf$ in \eqref{eq:f_Y(y) Mixtures}.
Equation \eqref{eq:k_y Mixt.Def.} suggests that in this case $k_y $ can be interpreted as a
weighted sum of other $L+1$ regression coefficients
\begin{equation}
\label{eq:k_y(l)Definition}
k_y^{(l)} =\frac{E_{Y_l } \{g(Y_l )Y_l \}}{\sigma _{Y,l}^2 },
\end{equation}
as expressed by
\begin{equation}
\label{eq:k_y_by_k_y(l)}
k_y =\sum\limits_{l=0}^L {\frac{\sigma _{Y,l}^2 }{\sigma _Y^2 }\beta _l
k_y^{(l)} } .
\end{equation}
Each gain $k_y^{(l)} $ in \eqref{eq:k_y_by_k_y(l)} is associated to the \textit{virtual} output $Z_l =g(Y_l )$, generated by the non-linearity $g(\cdot )$ when it is applied to the
Gaussian-distributed \textit{virtual} input $Y_l$.
Analogously

\begin{IEEEeqnarray}{ll}
\label{eq:k_x_by_k_x(l)}
k_x \; &=\frac{1}{\sigma_X^2}E_{XN} \{g(X+N)X\} = \sum\limits_{l=0}^L {\beta _l k_x^{(l)} }, \\
\label{eq:k_n_by_k_n(l)}
k_n \; &=\frac{1}{\sigma _N^2}E_{XN} \{g(X+N)N\} = \sum\limits_{l=0}^L {\frac{\sigma_{N,l}^2 }{\sigma_N^2 }\beta_l k_n^{(l)} },
\end{IEEEeqnarray}
where $k_x^{(l)}$ (and similarly $k_n^{(l)}$) is expressed by
\begin{equation}
\label{eq30}
k_x^{(l)} =\frac{E_{XN_l } \{g(X+N_l )X\}}{\sigma _X^2 }.
\end{equation}
Due to the fact that $X$, $N_l$, and $Y_l =X+N_l $, satisfy the hypotheses
of Theorem \ref{Th:EqGain-Gauss}, it is possible to conclude that
\begin{equation}
\label{eq:k_x(l)=k_y(l) Mixtures}
k_x^{(l)} =k_y^{(l)} =k_n^{(l)},
\end{equation}
which plugged in \eqref{eq:k_y_by_k_y(l)} leads to
\begin{equation}
\label{eq:k_y_by_k_x(l)}
k_y =\sum\limits_{l=0}^L {\frac{\sigma _{Y,l}^2 }{\sigma _Y^2 }\beta _l
k_x^{(l)} } .
\end{equation}
By direct inspection of \eqref{eq:k_y_by_k_x(l)}, \eqref{eq:k_x_by_k_x(l)}, and \eqref{eq:k_n_by_k_n(l)}, it is possible to conclude that $k_y
\ne k_x \ne k_n \ne k_y $, as soon as $L>0$, for any value of the weights $\beta _l $, and
any NLT $g(\cdot )$.
However, plugging \eqref{eq:k_x(l)=k_y(l) Mixtures} in \eqref{eq:k_x_by_k_x(l)}-\eqref{eq:k_n_by_k_n(l)}, it is obtained
\begin{IEEEeqnarray}{ll}
\label{eq:k_x_by_k_y(l)}
k_x = \sum\limits_{l=0}^L {\beta _l k_y^{(l)}}, \qquad \qquad
k_n = \sum\limits_{l=0}^L {\frac{\sigma_{N,l}^2}{\sigma_N^2} \beta _l k_y^{(l)}},
\end{IEEEeqnarray}
which may be considered the \emph{generalization} of \eqref{eq:k_y=k_x=k_n}, when $X$ is a zero-mean Gaussian and $N$ a zero-mean Gaussian-mixture. Indeed, also in this case the first equation in \eqref{eq:k_x_by_k_y(l)} is much simpler to compute than \eqref{eq:k_x_by_k_x(l)}, and enables the derivation of some useful theoretical results in estimation and information theory, as detailed in the next Sections. Finally, when both $X$ and $N$ are zero-mean independent Gaussian-mixtures, with parameters $\left(\beta_l^{(x)},\sigma_{X,l}^2, L_x \right)$ and $\left(\beta_l^{(n)},\sigma_{N,l}^2, L_n \right)$, respectively, \eqref{eq:k_y_by_k_y(l)} and \eqref{eq:k_x_by_k_y(l)} can be further generalized to
\begin{IEEEeqnarray}{ll}
\label{eq:k_y_k_x_kn Multiple Mixtures}
k_y =\sum\limits_{l=0}^{L_x} {\sum\limits_{j=0}^{L_n} {\beta_l^{(x)} \beta_j^{(n)}\frac{\sigma_{Y,(l,j)}^2 }{\sigma_Y^2 } k_y^{(l,j)}}},\\
k_x =\sum\limits_{l=0}^{L_x} {\sum\limits_{j=0}^{L_n} {\beta_l^{(x)} \beta_j^{(n)}\frac{\sigma_{X,l}^2 }{\sigma_X^2 } k_x^{(l,j)}}}, \qquad
k_n =\sum\limits_{l=0}^{L_x} {\sum\limits_{j=0}^{L_n} {\beta_l^{(x)} \beta_j^{(n)}\frac{\sigma_{N,j}^2 }{\sigma_N^2 } k_n^{(l,j)}}},
\end{IEEEeqnarray}
where by intuitive notation equivalence, $Y_{l,j}= X_l + N_j$, $\sigma_{Y,(l,j)}^2 = \sigma_{X,l}^2 + \sigma_{N,j}^2$, $k_y^{(l,j)}=E\{g\left(Y_{l,j}\right)Y_{l,j}\}/\sigma_{Y,(l,j)}^2$, and $k_y^{(l,j)}=k_x^{(l,j)}=k_n^{(l,j)}$. Thus, also in this case, $k_y \ne k_x$, with the equality that is possible only if $X$ and $N$ are characterized by identical parameters $\left(\beta_l^{(o)},\sigma_{o,l}^2, L_o\right)$, e.g., if they are identical distributed, as envisaged by Theorem \ref{Th:EqGain-SamePDF}.

%
%
%

\section{INFORMATION AND ESTIMATION THEORETICAL IMPLICATIONS}
\label{sec:Inf&Est-Theory}
This section is dedicated to clarify how the theoretical results derived in Section \ref{sec:NL&Sum} and \ref{sec:MixtureExtension} are particularly pertinent to estimation and information theory, where Theorem \ref{Th:EqGain-Gauss} and its generalization in \eqref{eq:k_x(l)=k_y(l) Mixtures} find useful applications.
\begin{figure}[h]
\centerline{\includegraphics[width=\figwidth]{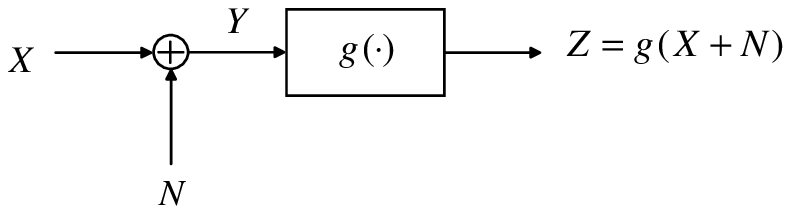}}
\caption{The statistical model}
\label{fig1}
\end{figure}
Indeed, it can be observed that the theoretical framework derived so far is captured by the model in
\figurename~\ref{fig1}, which is quite common for instance in several
communication systems, where $X$ may represent the useful information, $N$
the noise or interference, and $g(\cdot )$ either a distorting
non-linear device (such as an amplifier, a limiter, an analog-to-digital
converter, etc.), or an estimator/detector that is supposed to contrast the
detrimental effect of $N$ on $X$. Furthermore, the coefficient $k_y$ in \eqref{eq:Z=KyY}-\eqref{eq:k_y defin.} is the same coefficient that appears in the Bussgang theorem \cite{Bussgang:1952}, which lets to extend \eqref{eq:Z=KyY} to some special random processes, such as the Gaussian ones. Specifically, for the class of stationary Bussgang processes \cite{Nutall:1958},\cite{Rocca:1979}, it holds true that
\begin{equation}
\label{eq:Z(t)=KyX(t)+Wy(t)}
Z(t)=k_y Y(t)+W_y (t),
\end{equation}
where
\begin{equation}
\label{eq7}
k_y =\frac{R_{ZY}(0)}{R_{YY} (0)} =\frac{E\{Z(t)Y(t+\tau)\}}{E\{Y^2(t)\}} \qquad \qquad
,\forall t,\forall \tau,
\end{equation}
$R_{ZY} (\tau )=E\{Z(t)Y(t+\tau )\}$ is the classical
cross-correlation function for stationary random processes, and $R_{W_y Y}
\left(\tau \right)=0$, $\forall \tau$.
As detailed in Appendix \ref{App:Bussgang} the Bussgang theorem \cite{Bussgang:1952} can be exploited to prove Theorem \ref{Th:EqGain-Gauss}. Furthermore, it can also be used to characterize the power spectral density of the output of a non linearity
with Gaussian input processes. This fact induced an extensive technical
literature, with closed form solutions for the computation of the LRC $k_y$ for a
wide class of NLT $g(\cdot)$,
as detailed in \cite{Bussgang:1952, Baum:1957, Price:1958, Davenport:1958, Blachman:1968, Van:1966, Baum:1969, Levine:1973} for real Gaussian inputs, and in \cite{Minkoff:1985, Banelli:2000, Dardari:2000} for complex Gaussian inputs.
The Bussgang Theorem can also be used to asses the performance of such non-linear communication systems, such as the bit-error-rate (BER), the signal-to-noise power ratio $(\textrm{SNR})$, the maximal mutual information (capacity), and the mean square estimation error $(\textrm{MSE})$, whose
link has attracted considerable research efforts in the last decade (see \cite{Guo:2005, Prasad:2010} and references therein). Thus, taking in mind the broad framework encompassed by \figurename~\ref{fig1}, the following subsections will clarify how some of the theorems derived in this paper impact on the computations of the $\textrm{SNR}$, the capacity, and the $\textrm{MSE}$, and will provide also insights on their interplay in non-Gaussian and non-linear scenarios.


\subsection{SNR considerations}
In order to define a meaningful $\textrm{SNR}$, it is useful to separate the
non-linear device output as the sum of the useful information with an
uncorrelated distortion, as in (\ref{eq:Z=G(X+N)=KxX+Wx=KnN+Wn}).
For simplicity, we assume in the following that all the random variables are zero-mean, i.e., $P_X={\sigma_X}^2$. Thus, the $\textrm{SNR}$ at the non-linearity
output, is expressed by
\begin{IEEEeqnarray}{ll}
\label{eq:SNR_x}
 \textrm{SNR}_x \;&=k_x^2 \frac{E\{X^2\}}{E\{W_x^2 \}}=\frac{k_x^2 \sigma
_X^2 }{E\{Z^2\}-k_x^2 \sigma _X^2 }  \nonumber \\
 &=\left( {\frac{E_Y \{g^2(Y)\}}{k_x^2 \sigma _X^2 }-1} \right)^{-1},
\end{IEEEeqnarray}
where the second equality is granted by the orthogonality between $X$ and
$W_x$.

In the general case, in order to obtain a closed form expression for \eqref{eq:SNR_x},
it would be necessary to solve the double folded integral in \eqref{eq:k_x and k_n}, for the
computation of $k_x $. However, if $X$ and $N$ are zero-mean, independent, and
Gaussian, by Theorem \ref{Th:EqGain-Gauss} the computation can be simplified by exploiting
that $k_x =k_y$ and, consequently, the computation of the $\textrm{SNR}$ would
request to solve only single-folded integrals, e.g., \eqref{eq:k_y defin.} and $E_Y
\{g^2(Y)\}$. Note that, in this case also $Y=X+N$ would be Gaussian and,
consequently, the computations of $k_y$ and $E_Y
\{g^2(Y)\}$ can benefit of the results available in the literature \cite{Bussgang:1952, Baum:1957, Davenport:1958, Van:1966, Baum:1969, Levine:1973, Rowe:1982, Banelli:2000, Dardari:2000}.
\footnote{An alternative way to simplify the computation of the linear gain $k_x$ by a single-folded integral could exploit hybrid non-linear moments analysis of Gaussian inputs \cite{Cheded:1997} \cite{Scarano:1991}, where it is proven that $E\{Xg(Y)\}=E\{X[a_0+a_1(Y-E\{Y\})]\}$, with $a_0=E\{g(Y)\}$ and $a_1=E\{dg(Y)/dY\}$. When $Y=X+N$, with zero-mean $X$ and $N$, it leads to $E_{XN}\{xg(X+N)\}=\sigma_X^2 E_Y\{dg(Y)/dY\}$. This fact highlights that $k_y=k_x=E_Y\{dg(Y)/dY\}$, i.e., for Gaussian inputs the statistical linear gain $k_y$ is equivalent to the average of the first-order term of the MacLaurin expansion of the non linearity. Similarly, if $Y$ ($N$) is a Gaussian-Mixture, it is possible to exploit  $E\{Xg(Y)\}=\sum{}{}{\beta_lE\{Xg(Y_l)\}}$ and $E_{XY_l}\{Xg(Y_l)\}=\sigma_X^2 E_{Y_l}\{dg(Y)/dY\}$.}

Actually, it could be argued that the $\textrm{SNR}$ may be also defined by
exploiting \eqref{eq:Z=KyY} rather than \eqref{eq:Z=G(X+N)=KxX+Wx=KnN+Wn}. Indeed, by rewriting \eqref{eq:Z=KyY} as
\begin{equation}
\label{eq:Z=kyX+kyN+Wy}
Z=g(X+N)=k_y X+k_y N+W_y,
\end{equation}
it is possible to define another $\textrm{SNR}$, as expressed by

\begin{IEEEeqnarray}{ll}
\label{eq:SNR_y}
\textrm{SNR}_y \; &=\frac{k_y^2 E\{X^2\}}{k_y^2 E\{N^2\}+E\{W_y^2
\}}=\frac{k_y^2 \sigma _X^2 }{E_Y \{g^2(Y)\}-k_y^2 \sigma _X^2 } \nonumber \\
 &=\left( {\frac{E_Y \{g^2(Y)\}}{k_y^2 \sigma _X^2 }-1} \right)^{-1}.
\end{IEEEeqnarray}

Theorem \ref{Th:EqGain-Gauss} states that the two $\textrm{SNRs}$ in \eqref{eq:SNR_y} and \eqref{eq:SNR_x} are identical if $X$ and $N$ are zero-mean, independent, and Gaussian.
When $N$ (and/or $X$) is non-Gaussian, it is possible to approximate its \emph{pdf} with infinite accuracy [26] by the Gaussian-mixture \eqref{eq:f_N(n) Mixtures} in Section \ref{sec:MixtureExtension}, which represents a wide class of zero-mean noises with symmetrical $pdf\textrm{s}$. In this case, $k_x \ne k_y $ and \eqref{eq:SNR_x} should be used instead of \eqref{eq:SNR_y}. However, although \eqref{eq:SNR_y} cannot be used to compute the $\textrm{SNR}$, Theorem \ref{Th:EqGain-Gauss} turns out to be useful to compute $k_x$, by exploiting
\begin{equation}
\label{eq:k_x&k_x(l)SNR}
k_x =\sum\limits_{l=0}^L {\beta _l k_x^{(l)} }, \qquad \quad k_x^{(l)} =k_y^{(l)}
=\frac{E_{Y_l } \{g(Y_l )Y_l \}}{\sigma _{Y,l}^2 },
\end{equation}
which again involves only the computations of single-folded integrals.  Note that, all the integrals $E_{Y_l } \{g(Y_l )Y_l \}$ in \eqref{eq:k_x&k_x(l)SNR} share the same closed-form analytical solution for the Gaussian \emph{virtual} inputs $Y_l$.

\subsection{Estimation theory and MSE considerations}
\label{secMMSE}
The definition of the error at the non-linearity output may depend on the
non-linearity purpose. If the NLT $g(\cdot )$ represents an
estimator of $X$ given the observation $Y=X+N$, as expressed by
\begin{equation}
\label{eq:X_hat}
\hat {X}=g(X+N)=k_x X+W_x ,
\end{equation}
the estimation error is defined as
\begin{equation}
\label{eq:error_X_hat}
e=\hat {X}-X=(k_x -1)X+W_x .
\end{equation}
Exploiting the uncorrelation between $X$ and $N$, which induces
\begin{equation}
\label{eq:Wx_Power}
E\{W_{x}^2 \}=E_Y \{g^2(Y)\}-k_x^2 E\{X^2\},
\end{equation}
the $\textrm{MSE}$ at the non-linearity output can be expressed by
\begin{IEEEeqnarray}{ll}
\label{eq:MSE}
\textrm{MSE}=E\{e^2\}&=(k_x -1)^2E\{X^2\}+E\{W_{x}^2 \} \nonumber \\
 &=E_Y\{g^2(Y)\}+\left(1-2k_x\right)E\{X^2\}.
\end{IEEEeqnarray}
However, looking at \eqref{eq:X_hat} from another point of view, it is also possible to consider
$g(\cdot )$ as a distorting device that scales by $k_x $ the useful
information $X$, i.e, \eqref{eq:MSE} represents the MSE of a (conditionally) biased
estimator. In this view, it is possible to define an unbiased estimator
$\hat {X}_u =\hat {X}/k_x $ and the associated unbiased estimation error as
\begin{equation}
\label{eq:error_Unbiased}
e_u =\hat {X}/k_x -X=W_x /k_x ,
\end{equation}
whose mean square-value is expressed by
\begin{IEEEeqnarray} {ll}
\label{eq:MSE_unbiased}
\textrm{MSE}_u &=E\{e_u^2 \}=E\{W_{x}^2 \}/k_x^2 \nonumber \\
&= E_Y\{g^2(Y)\}/{k_x^2}-E\{X^2\} .
\end{IEEEeqnarray}
It is straightforward to verify that, for a given information power $E\{X^2\}$, the non-linearities that maximize the two $\textrm{MSE}$ are different, as expressed by
\begin{equation}
\label{eq:g_mmse}
\begin{array}{ll}
 g_{\textrm{mmse}} (\cdot )&=\mathop {\arg \min }\limits_{g(\cdot )} \left[
\textrm{MSE} \right]=\mathop {\arg \min }\limits_{g(\cdot )} \left[ {\log (\textrm{MSE})}
\right] \\
 &=\mathop {\arg \min }\limits_{g(\cdot )} \left[ {E\{g^2(Y)\}/k_x } \right],
\\
 \end{array}
\end{equation}
and
\begin{equation}
\label{eq:g_u-mmse}
g_{\textrm{u-mmse}} (\cdot )=\mathop {\arg \min }\limits_{g(\cdot )} \left[
{\textrm{MSE}_u } \right]=\mathop {\arg \min }\limits_{g(\cdot )} \left[
{E\{g^2(Y)\}/k_x^2 } \right].
\end{equation}
The first criterion corresponds to the classical Bayesian minimum MSE
(MMSE) estimator, that is $g_{\textrm{mmse}} (Y)=E_{X\vert Y} \{X\}$. By
means of \eqref{eq:SNR_x} and \eqref{eq:g_u-mmse}, the second criterion, which is the unbiased-MMSE
(U-MMSE) estimator, is equivalent to the maximum-$\textrm{SNR}$ (MSNR) criterion. Note
that $k_x $ depends on $g(\cdot )$ by \eqref{eq:k_x and k_n} and consequently, in general
\begin{equation}
\label{eq:g_u-mmse_approx}
g_{\textrm{u-mmse}} (\cdot )\ne \frac{g_{\textrm{mmse}} (\cdot
)}{k^\textrm{(mmse)}_x}.
\end{equation}
Indeed, the right-hand term in \eqref{eq:g_u-mmse_approx} is a (conditionally) unbiased estimator, but not the (U-MMSE) optimal one, because it has been obtained by first optimizing the MSE, and
by successively compensating the biasing gain $k_x$, while $g_{\textrm{u-mmse}}(Y)$
should be obtained the other way around, as expressed by \eqref{eq:error_Unbiased} and \eqref{eq:g_u-mmse}. The two criteria tend to be quite similar when the functional derivative
$\frac{\delta k_x (g(\cdot ))}{\delta g(\cdot )}\approx 0$ in the
neighborhood of the optimal solution $g_{\textrm{mmse}}(\cdot)$.

Actually, the MMSE and the MSNR criteria are equivalent from an information
theoretic point of view only when $g(\cdot )$ is linear, as detailed in
\cite{Guo:2005}, in which case $g_{\textrm{u-mmse}} (\cdot )$ is equivalent to right-hand
 side of \eqref{eq:g_u-mmse_approx}. For instance, this happens when $X$
and $N$ are both zero-mean, independent, and Gaussian as in Theorem \ref{Th:EqGain-Gauss}, in which case it is well known that \cite{Kay:1993}
\begin{equation}
\label{eq:g_L-mmse}
\hat {X}_{\textrm{mmse}} =g_{\textrm{mmse}} \left( Y \right)=\frac{\sigma _X^2
}{\sigma _X^2 +\sigma _N^2 }Y=\frac{\sigma _X^2 }{\sigma _X^2 +\sigma _N^2
}\left( {X+N} \right)
\end{equation}
is just a scaled version of the U-MMSE
\begin{equation}
\label{eq:g_L-U-mmse}
\hat {X}_{\textrm{u-mmse}} =g_{\textrm{u-mmse}} \left( Y \right)=Y=X+N.
\end{equation}
By noting that the $\textrm{SNR}$ is not influenced by a scaling coefficient, because
it affects both the useful information and the noise, it is confirmed that
for linear $g(\cdot )$ the MMSE optimal solution is also MSNR optimal
\cite{Guo:2005}.

Conversely, when $N$ is not Gaussian distributed, its $pdf$ may be (or approximated by) a Gaussian-mixture as in \eqref{eq:f_N(n) Mixtures}. In this case, analogously to
the consideration for the $\textrm{SNR}$ computation, Theorem \ref{Th:EqGain-Gauss} turns out to be useful to compute $k_x $, and thus the $\textrm{MSE}$ in \eqref{eq:MSE}, and \eqref{eq:MSE_unbiased}, by the single-folded integrals involved in \eqref{eq:k_x_by_k_y(l)}, rather than by the double-folded integrals in \eqref{eq:k_x_by_k_x(l)}. The reader interested in this point, may find a deeper insights and a practical application in \cite{Banelli:2011}, where these considerations have been fully exploited to characterize the performance of MMSE and MSNR estimators for a Gaussian source impaired by impulsive Middleton's Class-A noise.

\subsection{Capacity considerations}
  Equations \eqref{eq:Z=G(X+N)=KxX+Wx=KnN+Wn} or \eqref{eq:Z=kyX+kyN+Wy} can also be exploited to compute the mutual information of the non-linear information channel $X\to Z=g(X+N)$ summarized by \figurename~\ref{fig1}. Actually, the exact computation of the mutual information is in general prohibitive due to the complicated expression for the $pdf$ of the two disturbance components $W_x$ and $k_yN+W_y$, in \eqref{eq:Z=G(X+N)=KxX+Wx=KnN+Wn} and \eqref{eq:Z=kyX+kyN+Wy}, respectively. Anyway, it is possible to exploit the theoretical results derived so far, to establish some useful bounds on the mutual information in a couple of scenarios, as detailed in the following.
\vspace{6pt}
 \subsubsection{Non-linear channels with non-Gaussian noise}\hspace{0pt}\\
   When the noise $N$ is not Gaussian, it is difficult to compute in closed form the mutual information $I(X\to Y)$ even in the absence of the non-linearity $g(\cdot )$, and only bounds are in general available \cite{Ihara:1978}. Actually, when the noise $N$ is the Gaussian-mixture summarized by \eqref{eq:f_N(n) Mixtures}, it does not either exist a closed form expression for the differential entropy $h(N)$, which can only be bounded as suggested in \cite{Huber:2008}.
  However, when $X$ is Gaussian, the results in this paper can be exploited to compute simple lower-bounds for the mutual information $I(X,Z)$ at the output of any non linearity $Z=g(Y)$, which may model for instance A/D converters, amplifiers, and so forth. These lower bounds are provided by the AWGN capacity of \eqref{eq:Z=G(X+N)=KxX+Wx=KnN+Wn} and \eqref{eq:Z=kyX+kyN+Wy}, when the disturbance is modeled as (the maximum-entropy \cite{Diggavi:2001})
  zero-mean Gaussian noise with variance $E\left\{{Z^2}
  \right\}-k_x^2 \sigma _X^2 $ and $E\left\{ {Z^2} \right\}-k_y^2 \sigma _X^2
  $, respectively.
  Thus, exploiting (\ref{eq:Z=G(X+N)=KxX+Wx=KnN+Wn}) and (\ref{eq:SNR_x}), it is possible to conclude that
  \begin{equation}
  \label{eq:I(X,Z)_LowBound_X}
  I(X,Z)\ge C_{g(\cdot)}^{(\textrm{snr}_x)} =\frac{1}{2} \log (1+\textrm{SNR}_x ),
  \end{equation}
  while, by exploiting (\ref{eq:Z=kyX+kyN+Wy}) and (\ref{eq:SNR_y}), it would be possible to conclude that
  \begin{equation}
  \label{eq:I(X,Z)_LowBound_Y}
  I(X,Z)\ge C_{g(\cdot)}^{(\textrm{snr}_y)} =\frac{1}{2} \log (1+\textrm{SNR}_y ).
  \end{equation}
  By Theorem \ref{Th:EqGain-Gauss}, the two lower-bounds are equivalent if $X$ and $N$ are
  zero-mean independent Gaussians. Otherwise, the correct $\textrm{SNR}$ is \eqref{eq:SNR_x} and
  the correct lower bound is \eqref{eq:I(X,Z)_LowBound_X}. For instance, in the simulation examples either when $N$ is Laplace distributed and independent of $X$ (see \figurename~\ref{fig4}), or when it is Gaussian distributed and positively correlated with $X$ (see \figurename~\ref{fig5}), $k_x >k_y$ and   consequently by \eqref{eq:SNR_x} and \eqref{eq:SNR_y}, $C_{g(\cdot)}^{(\textrm{snr}_x)} >C_{g(\cdot)}^{(\textrm{snr}_y)}$.
  As detailed in the previous subsections, the computations of such lower bounds are simplified by the results in this paper when $X$ is zero-mean Gaussian, and $N$ is either zero-mean Gaussian or a Gaussian mixture.

\vspace{6pt}
\subsubsection{Linear channels with non-Gaussian noise}\hspace{0pt}\\
It is also possible to derive a bound for the mutual information of the non-Gaussian additive channel $Y=X+N$, in the absence or before the NLT $g(\cdot)$, by exploiting the interplay between $\textrm{MSE}$ and mutual information. Indeed, for non-Gaussian additive channels, exploiting the corollary of Theorem 8.6.6 in \cite{Cover:2006}, it is possible to readily derive that
 \begin{equation}
  \label{eq:MutInf_LowerBound}
     I(X,Y)\ge h(X) - \frac{1}{2}\log \left(2\pi e \; \textrm{MSE} \right).
  \end{equation}
 which holds true for the $\textrm{MSE}$ of any estimator $\hat{X}=g(Y)$.
 Thus, for a Gaussian source $X$, \eqref{eq:MutInf_LowerBound} simply becomes
\begin{equation}
\label{eq:I(X,Y)_MSE_LB}
  I(X,Y)\ge C_{g(\cdot)}^{\textrm{(mse)}} = \frac{1}{2}\log\left(\frac{\sigma_{X}^2}{\textrm{MSE}}\right),
  \end{equation}
  where, the lower bound $C_{g(\cdot)}^{\textrm{(mse)}}$ can be computed by plugging \eqref{eq:MSE} in \eqref{eq:I(X,Y)_MSE_LB}. Taking in mind that an estimator is generally non-linear, it is possible to exploit the information processing inequality \cite{Cover:2006},
to establish another lower bound by means of \eqref{eq:I(X,Z)_LowBound_X}
\begin{equation}
\label{eq:I(X,Y)_InfProcIneq}
I(X,Y) \ge I(X,\hat{X}(Y))
\ge  C_{g_{\textrm{mmse}}(\cdot)}^{(\textrm{snr}_x)},
\end{equation}
by properly computing the linear gain $k_x$ and output power $E\{\hat{X}(y)^2\}$ associated to the estimator $\hat{X}(Y)$.
It is natural to ask which of the two bounds in \eqref{eq:I(X,Y)_MSE_LB} and \eqref{eq:I(X,Y)_InfProcIneq} is the tightest, and should be used in practice. To this end, lets note that by \eqref{eq:SNR_x} and \eqref{eq:MSE} $\textrm{MSE}$ and $\textrm{SNR}_x$ are linked by
  \begin{equation}
  \label{eq:SNRvsMSE}
  \textrm{SNR}_x= \frac{k_x^2\sigma_X^2}{E\{W_x^2\}} = \frac{k_x^2\sigma_X^2}{\textrm{MSE}-\left(1-k_x\right)^2\sigma_X^2},
\end{equation}
which lets to establish the following general Theorem
\vspace{5mm}
\begin{theorem}
\label{Th:CapacityBounds}
\textit{For any additive noise channel $Y=X+N$, and any estimator $\hat{X}(Y)$, the capacity lower bound based on the} SNR \textit{is always tighter, (or at least equivalent), than the capacity lower bound based on the} MSE\textit{, as summarized by}
\begin{equation}
\label{eq:CapacityBounds}
C_{g(\cdot)}^{(\textrm{snr}_x)} \ge C_{g(\cdot)}^{(\textrm{mse})}.
\end{equation}
\end{theorem}
\begin{IEEEproof}
See Appendix \ref{App:CapacityBounds}.
\end{IEEEproof}
\vspace{5mm}
The two lower bounds are a valuable alternative to the pessimistic lower bound that models the noise as completely Gaussian, which is expressed by
\begin{equation}
\label{eq:I(X,Z)_LowBound_AWGN}
I(X,Y)\ge C_{\textrm{AWGN}} =\frac{1}{2} \log (1+\textrm{SNR}),
\end{equation}
where the total SNR is defined as $\textrm{SNR}=\frac{\sigma_x^2}{\sigma_n^2}$.  For any estimator such that $\textrm{MSE} \le \sigma_n^2 \frac{SNR}{SNR+1}$, by means of \eqref{eq:I(X,Y)_MSE_LB} and \eqref{eq:I(X,Z)_LowBound_AWGN}, $C_{g(\cdot)}^{(\textrm{mse})} \ge C_{\textrm{AWGN}}$. Actually, any useful estimator should significantly reduce the estimation error power with respect to the original noise power [e.g., the estimation error power with $g(y)=y$], as expressed by $\textrm{MSE} \ll \sigma_n^2$: this fact consequently induces that $C_{g(\cdot)}^{(\textrm{mse})}>C_{\textrm{AWGN}}$ is verified for any practical estimator and $\textrm{SNR}$, as it will be confirmed in the simulations section.
Note that, the lower bound in \eqref{eq:MutInf_LowerBound} has been also derived in \cite{Prasad:2010} for the MMSE estimator $g_{\textrm{mmse}}(\cdot)$, which obviously provides the tightest $\textrm{MSE}$ bound among all the estimators. In the light of Theorem \ref{Th:CapacityBounds}, the bound in \eqref{eq:I(X,Z)_LowBound_X} together with \eqref{eq:SNRvsMSE} is an alternative (possibly better) approximation of the relationship between mutual information and $\textrm{MMSE}$, which recently attracted several research \cite{Guo:2005} \cite{Prasad:2010}.

Applying the analytical framework derived in this paper, the general result given by Theorem \ref{Th:CapacityBounds}, can be exploited when the noise $N$ can be modeled, or approximated, by the Gaussian-mixture in \eqref{eq:f_N(n) Mixtures}, as in the case of a Class-A impulsive noise. Indeed, in this case Theorem \ref{Th:EqGain-Gauss} turns out to be useful to establish both the $\textrm{MSE}$ bound in \eqref{eq:I(X,Y)_MSE_LB}, and the tighter bound $C_{g(\cdot)}^{(\textrm{snr}_x)}$ in \eqref{eq:I(X,Z)_LowBound_X} because, as already explained, the computation of the gain $k_x$ in \eqref{eq:k_x&k_x(l)SNR} and $E\{g^2(Y)\}$ involve only single-folded integrals. The tightest bounds would be provided by the MMSE estimator, i.e., by computing \eqref{eq:k_x&k_x(l)SNR} and $E\{g^2(Y)\}$ with  $g(\cdot)=g_{\textrm{mmse}}(\cdot)$: actually, for a Gaussian-mixture noise the MMSE estimator is characterized by the rather involved expression \cite{Banelli:2011}
\begin{equation}
\label{eq:OBE_final}
g_{\textrm{mmse}}(y)
=\frac{\sum\limits_{m=0}^\infty
{\frac{\sigma_X^2}{\sigma_X^2 +\sigma _m^2 }\beta _m G(y;\sigma_X^2
+\sigma _m^2 )} }{\sum\limits_{m=0}^\infty {\beta _m G(y;\sigma_X^2 +\sigma
_m^2 )} }y,
\end{equation}
which prevents closed form solutions. Thus, the computation of the lower bound in \eqref{eq:I(X,Y)_MSE_LB} requests (single-folded) numerical (or Montecarlo) integration techniques\footnote{An alternative numerical approach to the computation of $E\{g_{\textrm{mmse}}^2(Y)\}$ is to expand $g_{\textrm{mmse}}(\cdot)$ as a series of opportune functions (Hermite polynomials, etc.) that admit closed form expressions for their averages over Gaussian $pdf$s (see \cite{Davenport:1958}, \cite{Levine:1973}, \cite{Banelli:2000} and references therein). This is however out of the scope of this paper, and a possible subject for further investigations.}. Alternatively, in order to come up with capacity lower bounds (e.g., $\textrm{MSE}$ and $\textrm{SNR}_x$) in closed form expressions, it is possible to exploit a suboptimal estimator for the Class-A noise, such as the blanker non-linearity (BN)
\begin{equation}
\label{eq:g_BN}
g_{\textrm{BN}}(y)= y \cdot \textrm{u}_{\textrm{-1}}\left(y_{\textrm{th}}-|y|\right),
\end{equation}
which nulls out all the inputs, whose absolute value overpasses a (MMSE optimal) threshold $y_{\textrm{th}}$ \cite{Banelli:2011} \cite{Zhidkov:2008}.
Such a BN is slightly suboptimal in MSE (and SNR) with respect to the MMSE estimator, and consequently provides slightly looser lower bounds with respect to $C_{g_{\textrm{mmse}}(\cdot)}^{(\textrm{snr}_x)}$ and $C_{g_{\textrm{mmse}}(\cdot)}^{\textrm{(mse)}}$, as it will be verified in the next section.



\section{SIMULATIONS}
\label{sec:Simulations}
This section reports some computer-aided simulations to give further evidence and insights to the Theorems, and also to assist the estimation and information theory implications.
To this end, it is considered a simple soft-limiting (SL) NLT
\begin{equation}
\label{eq33}
g_{\textrm{SL}}(y)=\left\{ {\begin{array}{ll}
 y	&,\vert y\vert <y_{\textrm{th}} \\
 y_{\textrm{th}} \textrm{sign}(y)	&,\vert y\vert \ge y_{\textrm{th}} \\
 \end{array}} \right..
\end{equation}
In a first set of simulations the clipping threshold has been fixed as $y_{\textrm{th}} =1$, and the average input power is always set to $P_Y =10$, in order to evidence the non-linear behavior,
by frequently clipping the input $Y=X+N$. Samples of the random variables
$X$ and $N$ have been generated according to either a zero-mean Gaussian
$\left[ \textrm{i.e., } f(\alpha )=G(\alpha ;\sigma ^2)\right]$, or a zero-mean Laplace
$pdf$ $\left[\textrm{i.e., } f(\alpha )=L(\alpha ;\sigma^2)=0.5\lambda e^{-\frac{\sqrt{2}\vert \alpha \vert}{\sigma} }\right]$, or a uniform $pdf$ $\left[\textrm{i.e., } f(\alpha )=U(\alpha ;\sigma^2)= \right.$
$\left. 0.5 \textrm{u}_{\textrm{-1}}\left(\vert \alpha-\sqrt{3}\sigma_x \vert \right)\right]$, or a triangular zero-mean $pdf$  $\left[\textrm{i.e., } f(\alpha )= \right.$ $\left.T(\alpha ;\sigma^2)=U(\alpha ;\sigma^2/2) \ast U(\alpha ;\sigma^2/2) \right]$.
The regression coefficients $k_y $, $k_x $, and $k_n $ have been estimated by
substituting each expected value in \eqref{eq:k_y defin.} and \eqref{eq:k_x and k_n}
, with the corresponding sample-mean over $10^6$ samples.

\begin{figure}[htbp]
\centering
\subfigure[$X\sim G(x;P_X )$, $N\sim G(n;P_N)$]
{\includegraphics[width=0.5\figwidth]{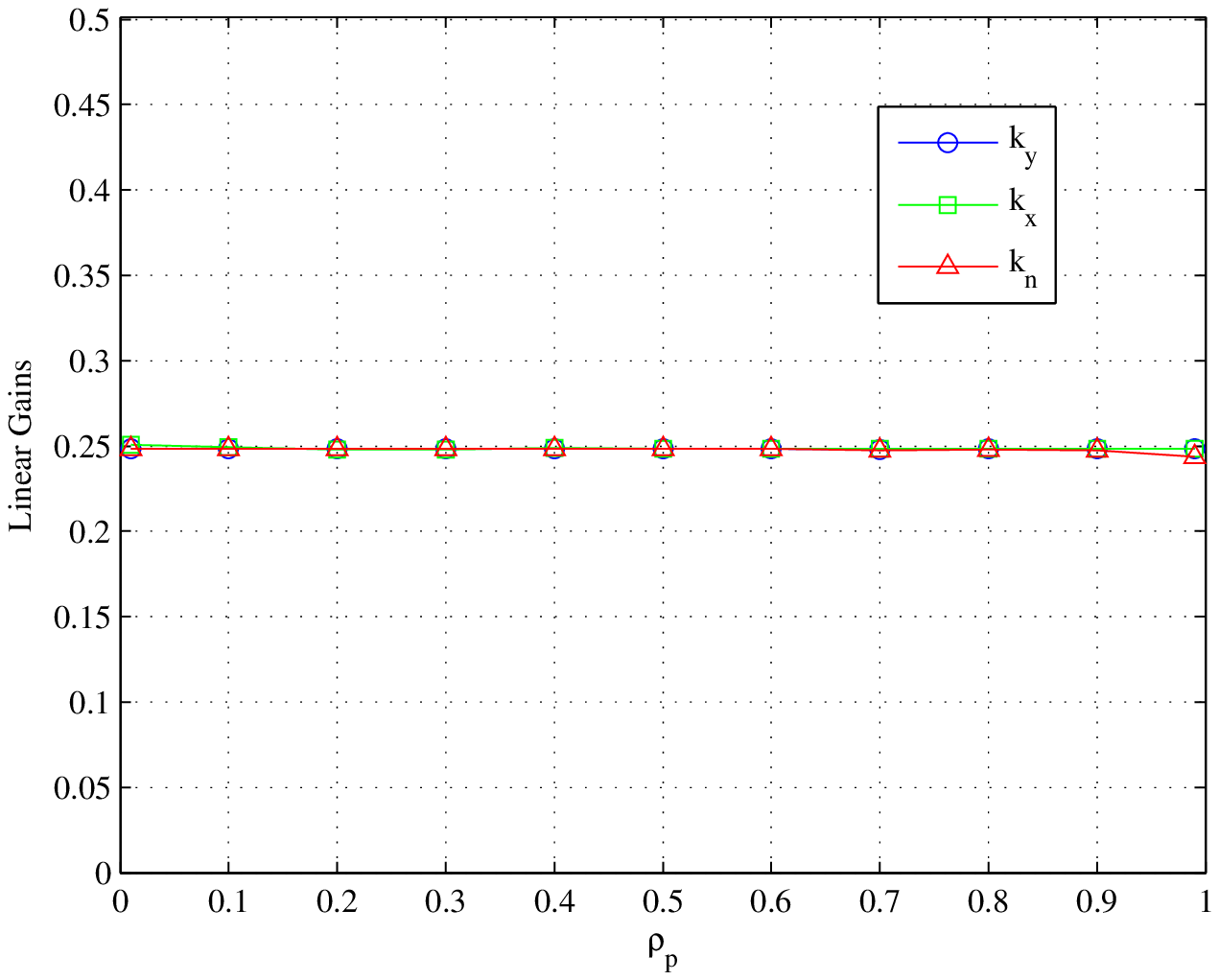}
\label{fig2}} \qquad
\subfigure[$X\sim L(x;P_X )$, $N\sim L(n;P_N)$]
{\includegraphics[width=0.5\figwidth]{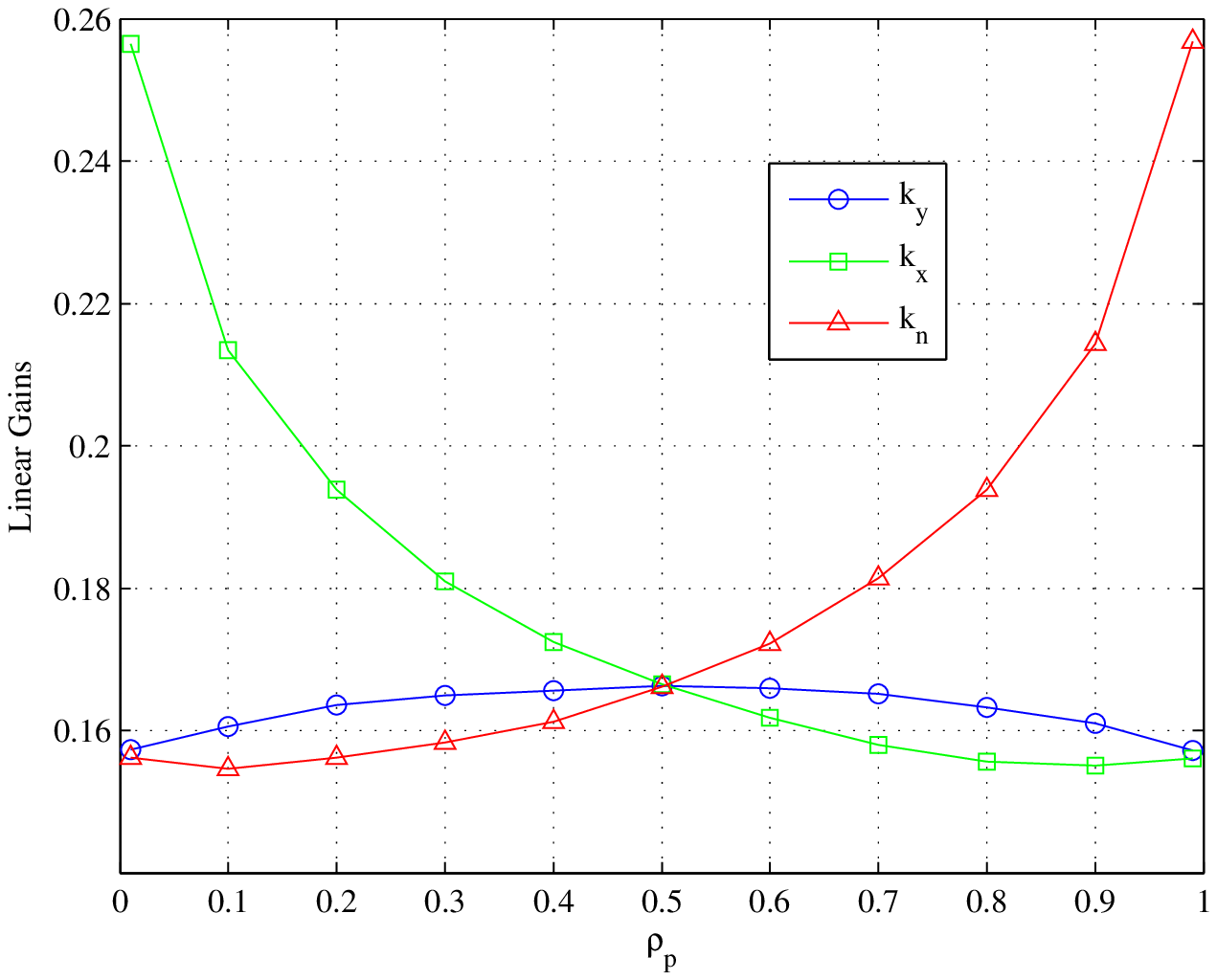}
\label{fig3}} \\
\subfigure[$X\sim G(x;P_X )$, $N\sim L(n;P_N)$]
{\includegraphics[width=0.5\figwidth]{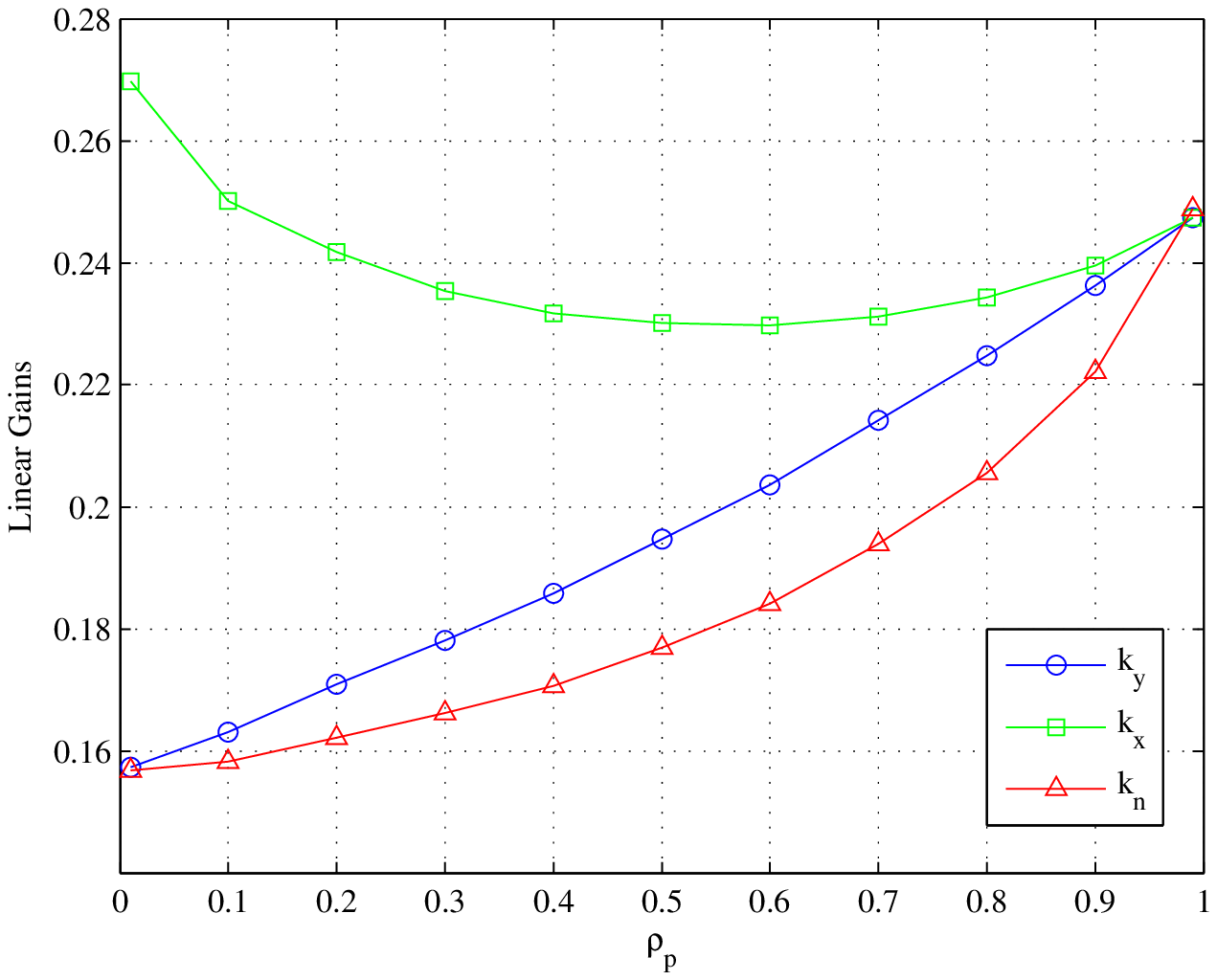}
\label{fig4}} \qquad
\subfigure[$X\sim G(x;P_X )$, $N\sim G(n;P_N)$, $\rho _{XN} =0.3$]
{\includegraphics[width=0.5\figwidth]{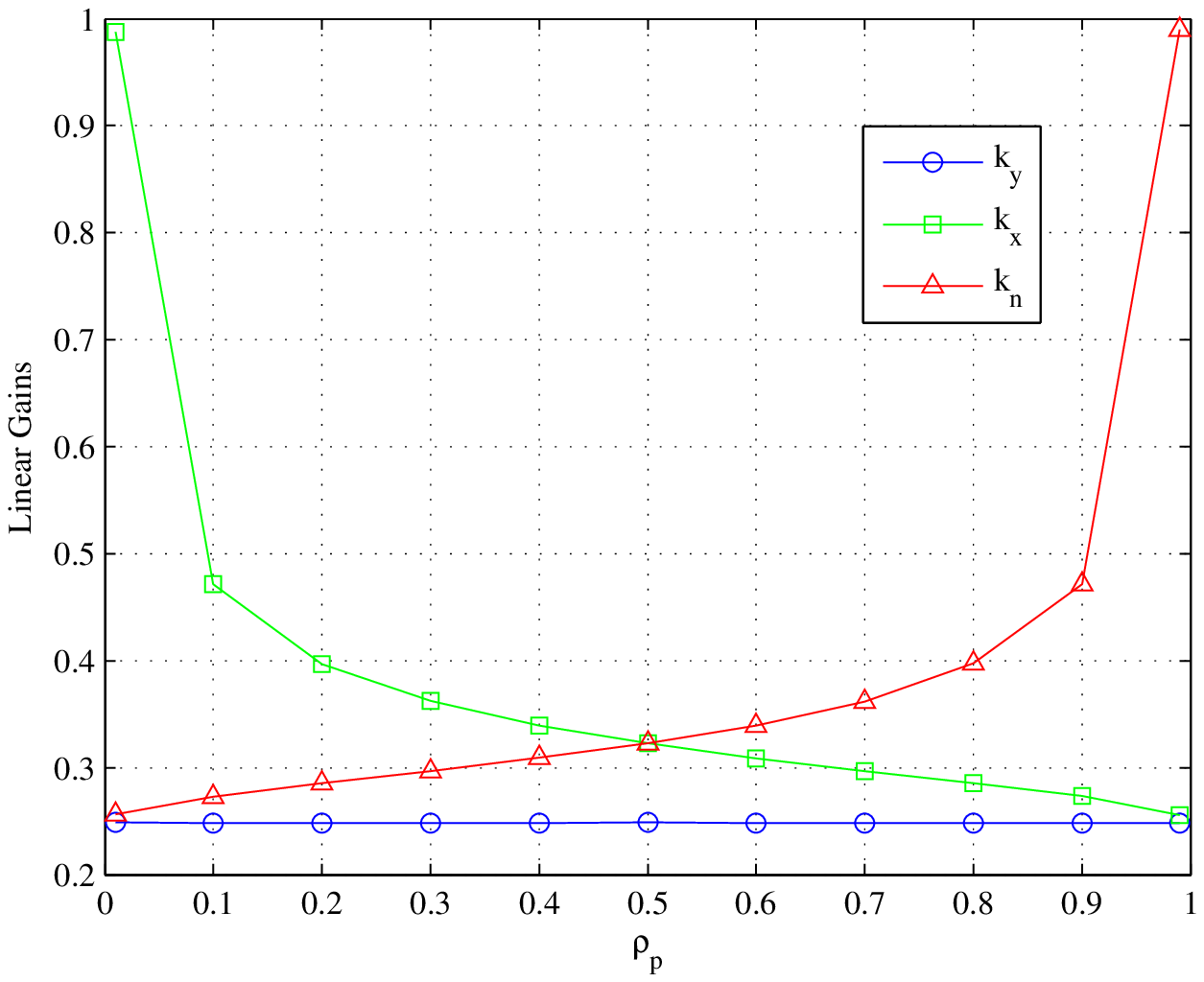}
\label{fig5}}
\caption{Linear regression coefficients versus the input power ratio, when $P_Y =10$ and the
inputs are a) independent and Gaussians {pdf}s; b) independent and Laplace {pdf}s; c) independent Gaussian and Laplace {pdf}s; d) correlated Gaussians {pdf}s.}
\label{Fig:Kx4}
\end{figure}

\figurename~\ref{fig2}-\figurename~\ref{fig5} plot the linear-regression coefficients versus
the mean square ratio $\rho _p =P_X /(P_X +P_N )$, which represents the
power percentage of $Y=X+N$ that is absorbed by $X$, when $X$ and $N$ are
independent.

\figurename~\ref{fig2}, where the input of the soft-limiter is the
sum of two independent zero-mean Gaussians, confirms Theorem \ref{Th:EqGain-Gauss},
with all the three regression coefficients that are identical, independently of
how the input power $P_Y =P_X +P_N $ is split between $X$ and $N$.


Conversely, in \figurename~\ref{fig3} the input is the sum of two (zero-mean)
independent Laplace random variables, and $k_y \ne k_x \ne k_n $. However,
when $\rho _p =0.5$, i.e., when the input power $P_Y $ is equally split
between $X$ and $N$, the three coefficients are equal, as predicted by Theorem
\ref{Th:EqGain-SamePDF}.


In \figurename~\ref{fig4}, where $X$ is zero-mean Gaussian while $N$ is an independent zero-mean Laplacian, it is clearly shown that $k_y \neq k_x \neq k_n$ for any $\rho_p$, as it happens in general.


This is also confirmed by \figurename~\ref{fig5} where, differently from \figurename~\ref{fig2}, the two Gaussian inputs $X$ and $N$ are not independent, and they are correlated with a correlation coefficient $\rho _{XN} =0.3$. It is observed that also in this case, all the regression coefficients are different, except when $\rho_p =0.5$, i.e., when $P_X =P_N $ and each variable absorbs a
fraction equal to $(1-2\rho _{XN} )/2$ of the total power $P_Y$.
Note however that, also in this specific case where $P_X =P_N $, $k_y <k_x =k_n$ due to (\ref{eq:ky-kx-kn general2}), which becomes $k_y =k_x/(1+\rho _{XN} )$.
Additionally, it is possible to observe that $k_y $ in
\figurename~\ref{fig5} should be equal to the value in \figurename~\ref{fig2},
because the non-linearity in both cases has a Gaussian input $Y$, with the
same power $P_Y =\sigma _Y^2 =10$. Another interpretation of this result is
the following: due to the correlation $\rho _{XN} $, it is possible to
express each separate component, for instance $N$, as a function of the other
one, i.e., $N=\rho _{XN} X+\varepsilon$, with $\varepsilon \sim
G(0,\sigma _\varepsilon ^2 )$, $\varepsilon $ independent of $X$, and
$\sigma _\varepsilon ^2 $ such that $P_Y =(1+\rho _{XN} )^2\sigma _X^2
+\sigma _\varepsilon ^2 $. Thus, for $Y=U+\varepsilon$, $U=(1+\rho_{XN})X$ the hypotheses of
Theorem \ref{Th:EqGain-Gauss} are satisfied and consequently $k_y =k_u =k_\varepsilon $, where
by straightforward substitutions $k_u =E\{ZU\}/P_U =k_x /(1+\rho _{XN} )$.

\begin{figure}[htbp]
\centering
\subfigure[$X\sim U(x;P_X)$, $N \sim U(n;P_N/2) \ast U(n;P_N/2)$]
{\includegraphics[width=0.5\figwidth]{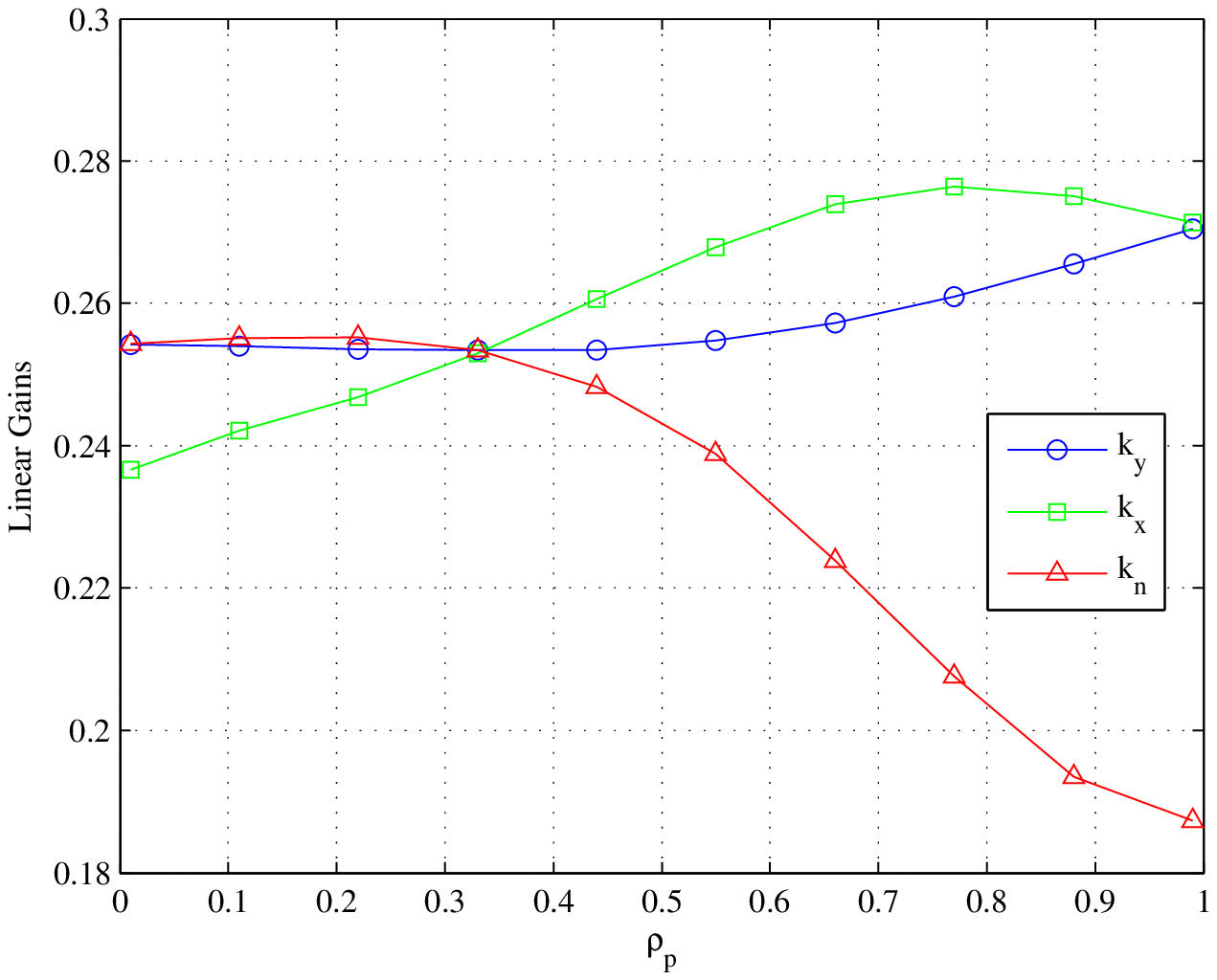}
\label{fig6}} \qquad
\subfigure[$X\sim N(x;P_X)$, $N \sim$ Middleton's Class-A noise]
{\includegraphics[width=0.5\figwidth]{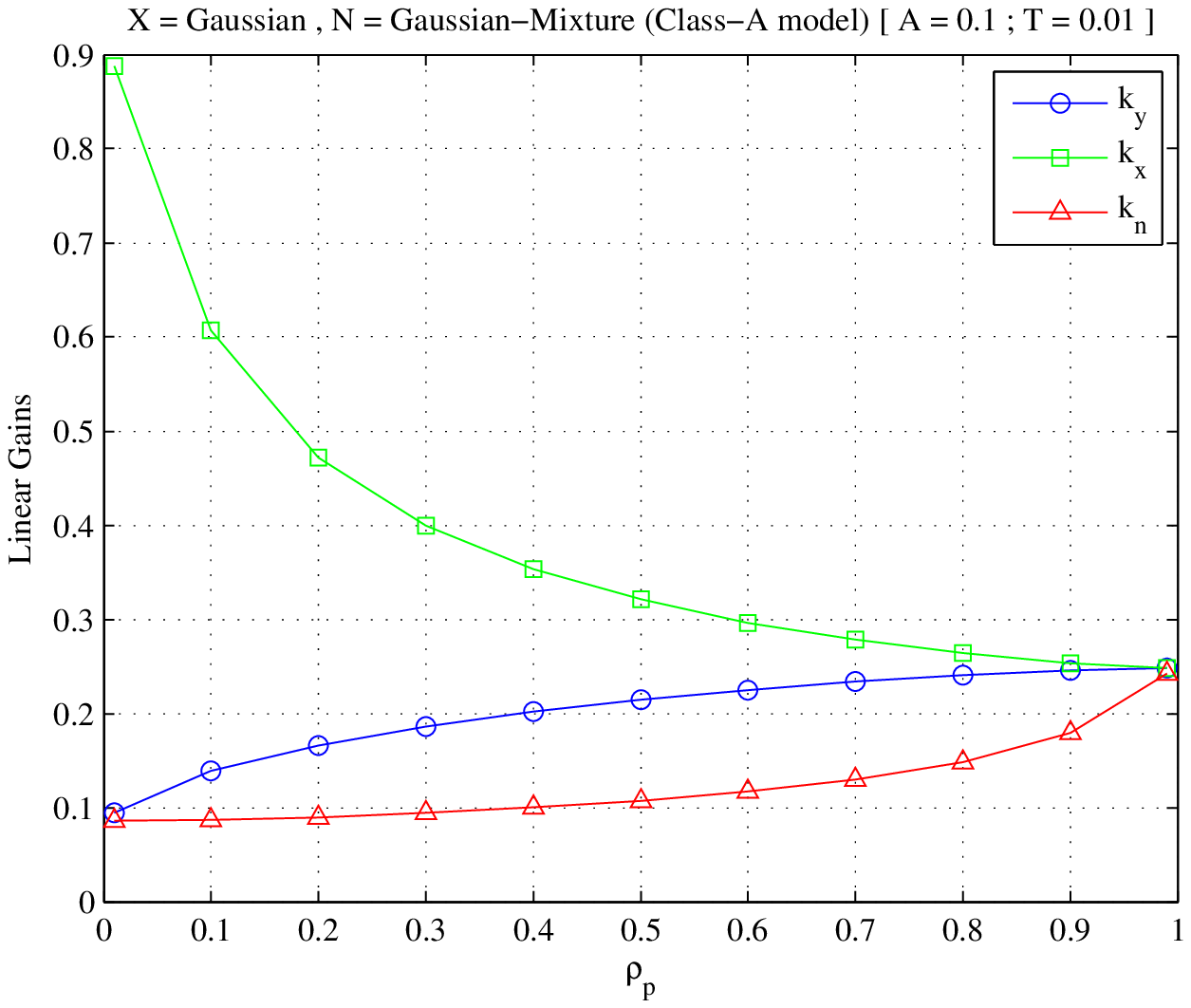}
\label{figNEW}}
\caption{Linear regression coefficients versus the input power ratio, when $P_Y =10$ and the
inputs are independent a) Uniform and Triangular {pdf}s; b) Gaussian and Gaussian-Mixture {pdf}s}
\label{Fig:Kx2}
\end{figure}


In \figurename~\ref{figNEW} $X \sim U(x;\sigma_X^2)$ is a zero-mean uniform random variable and $N \sim T(n;\sigma_N^2)$ has an independent zero-mean triangular \emph{pdf}: it can be observed that in general $k_y \ne k_x \ne k_n$ unless when $P_N=2P_X=2P_Y/3$ ($\rho_p=1/3$), i.e., when $f_N(n)=f_X(n)\ast f_X(n)$. This fact confirms Example \ref{Example:Convolutions} in Appendix \ref{App:EquivalenceTheorem}, where, generalizing Theorem \ref{Th:EqGain-SamePDF}, it has been highlighted that in this case $Y$ can be interpreted as $Y=X+(N1+N2)$, e.g., as the sum of three (uniform) i.i.d. random variables, and $k_y=k_x=k_{n_1}=k_{n_2}$.


A final set of results is dedicated to derive capacity bounds for a Gaussian source $X$ impaired by an impulsive noise $N$, modeled as a Gaussian mixture, according to the Middleton's Class-A noise model.
The analytical expression in \eqref{eq:OBE_final} has been used to compute by a Montecarlo semi-analytical approach $k_x^{\textrm{(mmse)}}=E\{xg_{\textrm{mmse}}(y)/\sigma_X^2\}$ and $E\{g_{\textrm{mmse}}(y)^2\}$: the obtained values are substituted in \eqref{eq:MSE} and \eqref{eq:SNR_x} to estimate the mutual information lower bounds in \eqref{eq:I(X,Y)_MSE_LB} and \eqref{eq:I(X,Z)_LowBound_X}, respectively.
\begin{figure}[h]
\centerline{\includegraphics[width=\figwidth]{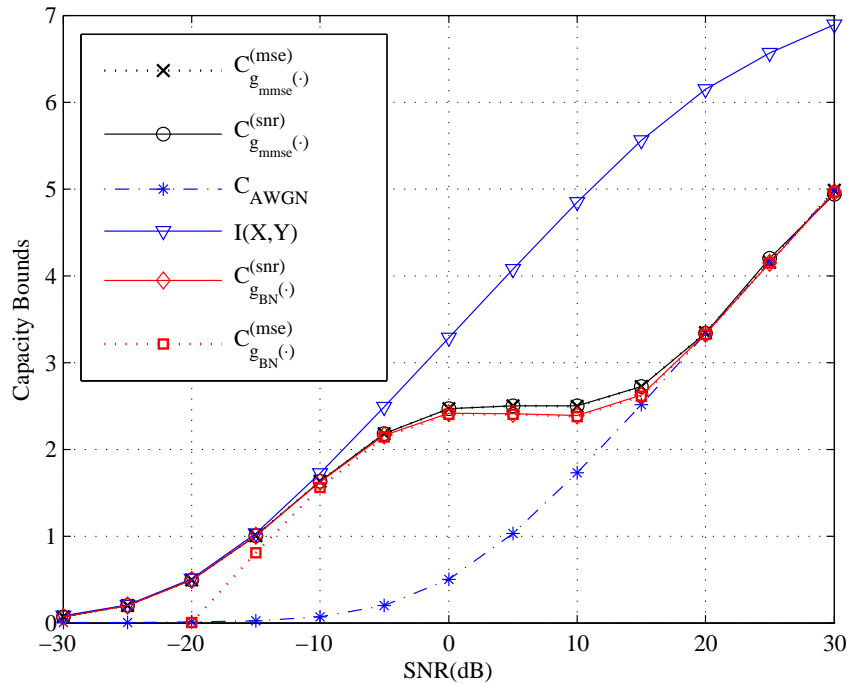}}
\caption{Capacity lower-bounds, for a zero-mean Gaussian source impaired by a Class-A (Gaussian-Mixture) impulsive noise with $A=0.01$ and $T=0.01$}
\label{Fig:CapacityBounds}
\end{figure}
Fig. \ref{Fig:CapacityBounds} shows the capacity bounds versus $\textrm{SNR}$ when the impulsive noise is characterized by the parameters $A=0.01$ and the power-ratio between $\textrm{AWGN}$ and impulsive noise $T=\sigma_t^2/\sigma_I^2=0.01$. Furthermore, Fig. \ref{Fig:CapacityBounds} shows also the mutual information $I(X,Y)$, which has been computed by approximating the joint and marginal $pdf$s of $X$ and $Y$ by the corresponding histograms, obtained by simulation trials over $10^8$ samples. It is possible to appreciate that the mutual information lower bounds are tight when the total $\textrm{SNR} < 0 \textrm{ dB}$, while they are quite loose for total $\textrm{SNR}\ge 10 \textrm{ dB}$, where they almost coincide with the classical AWGN capacity lower bound in \eqref{eq:I(X,Z)_LowBound_AWGN}. Note anyway that the total $\textrm{SNR}$ is defined as $\textrm{SNR}=\frac{\sigma_x^2}{\sigma_t^2+\sigma_I^2}=\frac{T}{T+1}\textrm{SNR}_{\textrm{awgn}}$, which in this case leads to $\textrm{SNR} \simeq 0.01\cdot \textrm{SNR}_{\textrm{awgn}}$: thus, the bounds are quite tight, and useful,  for $\textrm{SNR}_{\textrm{awgn}} \in [-10,20] \textrm{ dB}$, in the presence of strong impulsive noise, which is a regime of practical interest.
Fig. \ref{Fig:CapacityBounds} confirms that $C_{g(\cdot)}^{(\textrm{snr}_x)} \ge C_{g(\cdot)}^{(\textrm{mse})}$, as predicted by Theorem \ref{Th:CapacityBounds}: this is clearer at low $\textrm{SNR}$s for the suboptimal BN estimator $g_{\textrm{BN}}(\cdot)$ \cite{Banelli:2011} \cite {Zhidkov:2008}, which allow the closed form computations of the two lower bounds (e.g., of $k_x$ and $E\{g(\cdot)^2\}$). Conversely, the two lower bounds seem to coincide for the MMSE estimator $g_{\textrm{mmse}}(\cdot)$. Further note that the lower-bound $C_{g_{\textrm{BN}(\cdot)}}^{(\textrm{snr}_x)}$ is a tight approximation of the MMSE lower bound $C_{g_{\textrm{mmse}(\cdot)}}^{(\textrm{mse})}$.
A deeper analysis for different values of the Class-A noise parameters $A$ and $T$, as well as for different Gaussian mixture noises may be the subject for future investigation and is beyond the scope of this paper, whose aim is to establish the general theoretical framework.



\section{CONCLUSIONS}
\label{sec:conclusions}
The first contribution of this paper has been to prove and analyze some general and interesting theorems for non-linear transformations of the sum of independent Gaussian random variables, and Gaussian-Mixtures. Due to the widespread use of Gaussian and Gaussian-Mixtures, these theorems can be useful in several fields, which include estimation theory, information theory, and non-linear system characterization. Furthermore, the paper has highlighted that these theorems are particularly useful to compute the $\textrm{SNR}$, the $\textrm{MSE}$, and mutual information bounds associated with communication systems dealing with non-linear devices, and/or impaired by a Gaussian-mixture noise.

\appendices
\section{}
\label{App:EquivalenceTheorem}
\begin{theorem}
\label{Th:CharacteristicFunctions}
\textit{ Given two independent random variables $X$, $N$, and $Y=X+N$
\begin{equation}
\label{eq:CharFunctions}
E_{X|Y}\{X\}=\alpha y \iff C_X^{1-\alpha}(u) = C_N^{\alpha}(u).
\end{equation}
}
\end{theorem}
\begin{IEEEproof}
Observing that
\begin{equation}
\label{eq:E_X|Y_definition}
E_{X|Y}\{X\}=\int_{-\infty}^{+\infty}{x f_{X|Y}(x;y)dx}\\
= \frac{1}{f_Y(y)}\int_{-\infty}^{+\infty}{x f_{XY}(x,y)dx},
\end{equation}
clearly l.h.s. of \eqref{eq:E_X|Y_definition} holds true if and only if
\begin{equation}
\label{eq:E_X|Y_alpha}
\int_{-\infty}^{+\infty}{x f_{XY}(x,y)dx}= \alpha y f_Y(y).
\end{equation}
If $Y=X+N$, with $X$ independent of $N$, it is well known \cite{Papoulis:1991} that $f_{XY}(x,y)= f_X(x)f_{Y|X}(y;x)= f_X(x)f_{N}(y-x)$ and $f_Y(y)= f_X(y) \ast f_N(y)$,
where $\ast$ stands for the convolution integral operator.
Thus, \eqref{eq:E_X|Y_alpha} becomes
\begin{equation}
\label{eq:E_X|Y_convolution}
p(y) \ast f_N(y) = \alpha y \cdot \left[f_X(y) \ast f_N(y)\right],
\end{equation}
where $p(x)=xf_X(x)$.
By applying the inverse Fourier transform, \eqref{eq:E_X|Y_convolution} becomes
\begin{equation}
\label{eq:E_X|Y_Fourier}
P(u)C_N(u) = \frac{\alpha}{j2\pi} \frac{d}{du}[C_X(u)C_N(u)],
\end{equation}
where $P(u)=\frac{1}{j2\pi}\frac{d}{du}\left[C_X(u)\right]=\frac{1}{j2\pi}C_X^{'}(u)$,
and consequently
\begin{equation}
\label{eq:E_X|Y_Fourier-2}
C_X^{'}(u)C_N(u) = \alpha \left[C_X^{'}(u) C_N(u) + C_X(u)C_N^{'}(u) \right].
\end{equation}
The last equality is a differential equation, with separable variables, as expressed by
\begin{equation}
\label{eq:C(u) differential}
(1-\alpha)\frac{C_X^{'}(u)}{C_X(u)} = \alpha \frac{C_N^{'}(u)}{C_N(u)},
\end{equation}
which can be solved by direct integration, leading to
\begin{equation}
\label{eq:Cx(u)Cn(u) by integration}
(1-\alpha)log\left(C_X(u)\right) = \alpha log(C_N(u)) + C_o,
\end{equation}
where $C_o=0$ is imposed by the boundary conditions $C_X(0)=C_N(0)=1$.
Equation \eqref{eq:Cx(u)Cn(u) by integration} is equivalent to
\begin{equation}
\label{eq:Cx(u)Cn(u) solutions}
C_X^{1-\alpha}(u) = C_N^{\alpha}(u),
\end{equation}
which concludes the proof.
\end{IEEEproof}

It is possible to observe that, for a given $f_X(x)$ [or a given $f_N(n)$], \eqref{eq:Cx(u)Cn(u) solutions} and \eqref{eq:CharFunctions} do not always admit a solution $f_N(n)$ [or $f_X(x)$]. For a fixed \emph{pdf} $f_X(x)$, the existence of a solution is equivalent to
\begin{equation}
\label{eq:InverseFourier}
f_N(n)=\mathcal{F}^{-1}\{C_X^{\rho}(u)\},
\end{equation}
i.e., to the existence of the inverse Fourier transform of $C_X^{\rho}(u)$, where $\rho=\frac{1-\alpha}{\alpha}=\frac{P_Y-P_X}{P_X}$. 
To this end, it can be observed that $\forall \rho>0$ the function $C_X^{\rho}(u)$ preserves the conjugate symmetry of $C_X(u)=C_X^{\ast}(-u)$ and the unitary area of the $\emph{pdf}$ by $C_X(0)=1$. Moreover, if $\rho \in [0,1]$ and if $\int_{-\infty}^{+\infty}{|C_X(u)|du}<+\infty$, then also $\int_{-\infty}^{+\infty}{|C_X^{\rho}(u)|du}<+\infty$, which is a sufficient condition for the existence of the inverse Fourier transform.
Although it is beyond the scope of the paper to establish (if possible) all the possible conditions  where \eqref{eq:Cx(u)Cn(u) solutions} or \eqref{eq:InverseFourier} admit feasible solutions, it is highlighted that $\rho=\frac{P_N}{P_X}$ when $X$ and $N$ are independent, and consequently $\rho \in[0,1]$ when $P_X \ge P_N$. Furthermore, some examples are listed in the following to clarify the subject and identifying some specific cases of interest.
\\

\begin{example}
\label{Example:Convolutions}
If $\alpha=p/q<1$ with $p,q \in \mathbb{N}$, i.e., $\alpha \in \mathbb{Q}$, then \eqref{eq:Cx(u)Cn(u) solutions} is equivalent to
\begin{equation}
\label{eq:Cx(u)Cn(u) Rational}
C_X^{q-p}(u) = C_N^{p}(u).
\end{equation}
This means that for a fixed  $f_X(x)$, and a fixed $\alpha=p/q<1$, Theorem \ref{Th:CharacteristicFunctions} holds true if the random variable $N$ is characterized by a \emph{pdf} $f_N(n)$ that satisfies
\begin{equation}
\label{eq:fx(x)fn(n) convolution}
  f_N(n) \underbrace { \ast  \cdots  \ast }_{q - p - 1}f_N(n)
  = f_X(n) \underbrace { \ast  \cdots  \ast }_{p}f_X(n).
\end{equation}
Note that \eqref{eq:fx(x)fn(n) convolution} is a (multiple) auto-deconvolution problem in $f_N(n)$, which is well known to be ill-posed for several functions $h(n)=f_X(n) \underbrace { \ast  \cdots  \ast }_{p}f_X(n)$, even in the simple case $q-p=2$ where $f_N(n) \ast f_N(n)=h(n)$.
\end{example}
The problem admits a solution when $\alpha=2/3$ ($\rho=1/2$), where it boils down to $f_N(n)=f_X(n)\ast f_X(n)$. This means that $N$ can be thought as the sum of two other (independent) random variables $N=N_1+N_2$, each one with the same distribution of $X$. This is actually equivalent to a generalization of Theorem \ref{Th:EqGain-SamePDF} to the sum of three i.i.d. random variables. The generalization to the sum of $Q+1$ i.i.d. random variables is obtained for $\alpha=Q/(Q+1)$ ($\rho=1/Q$).
\\
\begin{example}
\label{Example:Gaussians}
If $X$ is Gaussian, with mean $m_X$ and variance $\sigma_X^2$, then \eqref{eq:Cx(u)Cn(u) solutions} (apparently) admits always a solution for any $\alpha \in [0,1]$, and would lead us to (erroneously) conclude that also $f_N(n)$ should be non-zero mean Gaussian. Indeed, the characteristic function of a Gaussian \emph{pdf} is a Gaussian function, and any (positive) exponential of a Gaussian function is still a Gaussian function. Thus, recalling that $C_X(u)=e^{-2\left(\pi \sigma_X u\right)^2+j2\pi m_X u}$, we would conclude that
\begin{equation}
\label{eq:f_N(n)-Gaussian}
f_N(n)=\mathcal{F}^{-1}\{C_X^{\rho}(u)\} =
\mathcal{F}^{-1}\{e^{-2\left(\pi \sqrt{\rho} \sigma_X u\right)^2+j2\pi \rho m_X u}\}
= G(n-\rho m_X; \rho \sigma_X^2),
\end{equation}
which holds true when $\rho >0$, i.e., when $\alpha \in [0,1]$ and $P_Y>P_X$. Actually, it should be observed that right-hand side of \eqref{eq:f_N(n)-Gaussian} implicitly contains the constraints $\sigma_N^2=\rho \sigma_X^2$, $m_N=\rho m_X$ that, by the definition of $\rho$, can be jointly satisfied \emph{iff} $m_X=m_N=0$, and $\forall \sigma_X, \forall \sigma_N$. Thus, the equal gain condition holds true for Gaussian inputs, only if they are zero-mean, as expressed by Theorem \ref{Th:EqGain-Gauss}.
\end{example}
\begin{example}
\label{Example:Identical pdf}
When $\alpha=1/2$, i.e., $\rho=1$, equation \eqref{eq:CharFunctions} boils down to the trivial case $C_X(u)=C_N(u)$, i.e., the sufficient condition for $k_y=k_x=k_n$  is satisfied if the independent random variables $X$ and $N$ are identically distributed (and zero-mean) with $f_X(\cdot)=f_N(\cdot)$. This is an alternative proof for Theorem \ref{Th:EqGain-SamePDF}.
\end{example}

\section{}
\label{App:Bussgang}
An alternative proof of Theorem \ref{Th:EqGain-Gauss} for Gaussians r.v. can exploit the Bussgang Theorem for jointly-Gaussian random processes  $x(t)$ and $y(t)$, which states that \cite{Bussgang:1952, Rowe:1982, Papoulis:1991}
\begin{equation}
\label{eq:Bussgang Theorem}
E\{x(t)g[y(t+\tau)]\}= \frac{E\{y(t)g[y(t+\tau)]\}}{\sigma_Y^2}E\{x(t)y(t+\tau)\}
\qquad , \forall \tau.
\end{equation}
Setting $X=x(t)$, $Y=y(t)$, and $\tau=0$, then \eqref{eq:Bussgang Theorem} easily leads to
\begin{equation}
\label{eq:k_x by Bussgang}
k_x=k_y\frac{E\{XY\}}{\sigma_X^2},
\end{equation}
which reduces to $k_x=k_y$ for $Y=X+N$, when $X$ and $N$ are zero-mean and independent (and Gaussian to let $Y$ be Gaussian).

Some Lemmas of Theorem \ref{Th:EqGain-Gauss} follow.
\vspace{5 mm}
\begin{lemma}
\label{lem:PropGain-Gauss}
\textit{If $X$ and $N$ are zero-mean Gaussian and independent, $Y=\alpha _x X+\alpha _n N$, with $\alpha _x ,\alpha _n \in {\rm R}$, then}
\[
\frac{E\{ZY\}}{\sigma _Y^2 }=\frac{1}{\alpha _x }\frac{E\{ZX\}}{\sigma _X^2
}=\frac{1}{\alpha _n }\frac{E\{ZN\}}{\sigma _N^2 }.
\]
\end{lemma}
\begin{IEEEproof}
By Theorem \ref{Th:EqGain-Gauss} with $\tilde {X}=\alpha _x X$ and $\tilde {N}=\alpha _n N$.
\end{IEEEproof}

\vspace{5 mm}
\begin{lemma}
\textit{If $\; Y=\sum\limits_{j=1}^J {\alpha _j X_j }, \; \alpha _j \in R$,  $X_j$ and $N$ are independent zero-mean Gaussian random variables, then}
\[
\frac{E\{ZY\}}{\sigma _Y^2 }=\frac{1}{\alpha _i }\frac{E\{ZX_i \}}{\sigma
_{X_i }^2 }	\qquad ,\forall i.
\]
\end{lemma}
\begin{IEEEproof}
By Theorem 1 and Lemma 1 with $X=\alpha _i X_i $ and $N=\sum\limits_{(j\ne
i)}^{}{\alpha _j X_j }.$
\end{IEEEproof}

\section{Proof of Theorem \ref{Th:CapacityBounds}}
\label{App:CapacityBounds}
Proving that $C_{\textrm{lb}}^{(\textrm{snr}_x)} > C_{\textrm{lb}}^{(\textrm{mse})}$ corresponds to prove that $1+\textrm{SNR}_x > \frac{\sigma_X^2}{\textrm{MSE}}$.
Thus, when $|k_x|\ge1$ it is straightforward to verify that
\begin{equation}
1+\frac{k_x^2\sigma_X^2}{P_{W_x}} = 1+\frac{k_x^2\sigma_X^2}{\textrm{MSE}-\left(1-k_x\right)^2\sigma_X^2} \ge 1+\frac{k_x^2\sigma_X^2}{\textrm{MSE}} > \frac{\sigma_X^2}{\textrm{MSE}}.
\label{eq:SNRvsMSE_Inequality_1}
\end{equation}
More generally the inequality $1+\textrm{SNR}_x > \frac{\sigma_X^2}{\textrm{MSE}}$ holds true when
\begin{equation}
\frac{P_{W_x}+k_x^2\sigma_X^2}{P_{W_x}} \ge \frac{\sigma_X^2}{P_{W_x}+\left(1-k_x\right)^2\sigma_X^2},
\end{equation}
that is when
\begin{equation}
P_{W_x}^2 + 2P_{W_x} k_x \left(1-k_x\right)\sigma_X^2+ \left(1-k_x\right)^2 k_x^2\sigma_X^2 \ge 0.
\label{eq:SNRvsMSE_Inequality_2}
\end{equation}
Clearly, \eqref{eq:SNRvsMSE_Inequality_2} holds true when $|k_x| \le 1$, which together with \eqref{eq:SNRvsMSE_Inequality_1} lets to conclude that the inequality holds true for $\forall k_x \in{R}$, concluding the proof.

\bibliographystyle{IEEEtran}

\bibliography{IEEEabrv,Bibliography_abbrev}

\end{document}